\documentclass{aastex62}

\graphicspath{{./}{figures/}}

\received{March 12, 2019}
\revised{June 21, 2019}
\accepted{July 22, 2019}
\submitjournal{AJ}

\shorttitle{Flare Recovery with Differential Evolution}
\shortauthors{Lawson et al.}

\begin{document}

\title{Identification of Stellar Flares Using Differential Evolution Template Optimization}
\author{Kellen D. Lawson}
\affiliation{Homer L. Dodge Department of Physics and Astronomy, University of Oklahoma, 440 W. Brooks Street, Norman, OK 73019, USA}
\author{John P. Wisniewski}
\affiliation{Homer L. Dodge Department of Physics and Astronomy, University of Oklahoma, 440 W. Brooks Street, Norman, OK 73019, USA}
\author{Eric C. Bellm}
\affiliation{Department of Astronomy, University of Washington, Box 351580, Seattle, WA 98195, USA}
\author{Adam F. Kowalski}
\affiliation{Astrophysical \& Planetary Sciences, University of Colorado, Boulder, CO 80309, USA}
\author[0000-0003-4401-0430]{David L. Shupe}
\affiliation{IPAC, California Institute of Technology, 1200 E. California
             Blvd, Pasadena, CA 91125, USA}

\begin{abstract}

We explore methods for the identification of stellar flare events in irregularly sampled data of ground-based time domain surveys. In particular, we describe a new technique for identifying flaring stars, which we have implemented in a publicly available Python module called ``PyVAN''. The approach uses the Differential Evolution algorithm to optimize parameters of empirically derived light-curve templates for different types of stars to fit a candidate light-curve. The difference of the likelihoods that these best-fit templates produced the observed data is then used to delineate targets that are well explained by a flare template but simultaneously poorly explained by templates of common contaminants. By testing on light-curves of known identity and morphology, we show that our technique is capable of recovering flaring status in 69\% of all light-curves containing a flare event above thresholds drawn to include \textless1\% of any contaminant population. By applying to Palomar Transient Factory data, we show consistency with prior samples of flaring stars, and identify a small selection of candidate flaring G-type stars for possible follow-up.

\end{abstract}


\section{Introduction} \label{sec:intro}

The arrival of large-scale recent, current, and near-future ground- and space-based time domain photometric surveys (All-Sky Automated Survey for SuperNovae (ASAS-SN) \citep{shappee2014}; Dark Energy Survey (DES) \citep{dark2016}, Evryscope \citep{law2015}; Large Synoptic Survey Telescope (LSST) \citep{ivezic2008}; MEarth \citep{nutzman2008}; Palomar Transient Factory (PTF) \citep{law2009}; Sloan Digital Sky Survey (SDSS) Stripe 82 \citep{abazajian2009}; Zwicky Transient Facility (ZTF) \citep{bellm2018}; Kepler \citep{borucki2010}; Transiting Exoplanet Survey Satellite (TESS) \citep{ricker2015}) is helping to transform many areas of stellar astrophysics.  
However, each of these surveys is driven by different primary science motivations, leading each to differ in their filter combinations, cadences, durations, photometric precision, and dynamic ranges.  The challenge for many domains of stellar astrophysics research is how to utilize the incredible wealth of data provided by these heterogeneous surveys.

Stellar flares are produced by magnetic reconnection events in stellar photospheres that release copious amounts of energy across X-ray, UV, optical, IR, and radio wavelength regimes over time-scales of seconds to days. Diagnosing the underlying stellar physics that drives these flares has historically relied on targeted observing campaigns on small numbers of individual objects (see e.g. \citealt{moffett1974, lacy1976, hawley1991, kowalski2010, fuhrmeister2011, hilton2011, kowalski2013, hawley2014, silverberg2016, loyd2018a, kowalski2018}), where the cadence, wavelength coverage, and/or duration of observations were optimized to constrain the detailed behavior of flares during their rise and decay phases. Large-scale time domain surveys offer the promise of studying different niches of flare physics, based on each survey's design.  Analysis of larger populations of flare stars using time domain surveys that utilize sparse, long duration observations at one or more filters (e.g. \citealt{kowalski2009, walkowicz2011, yang2017, schmidt2018}) lack sufficient cadence and wavelength coverage to diagnose the detailed behavior of rise and decay events, but are ideal for sampling the frequency of the highest energy, most infrequent events. Determining the frequency of these high energy flares is of particular interest for their potential impact on the habitability of exoplanets \citep{segura2010, loyd2018b, howard2018}. Analysis of large populations of high cadence, long duration observations in a single filter \citep{davenport2016} can better diagnose the ensemble frequency of moderate amplitude flares, including their rise and decay phase behavior, at the expense of lacking broad wavelength coverage to sample how the time-dependent behavior of the full stellar atmosphere to flare events.

Identifying and classifying the myriad of different variable sources detected in large volume time domain surveys is a subject of critical importance. The Arizona-NOAO Temporal Analysis and Response to Events System (ANTARES), for example, is currently developing and testing machine learning toolsets to characterize and classify variable and transient sources for LSST, to help populate data alert systems \citep{narayan2018}. \citet{zinn2017} have explored use of quasi-periodic oscillation stochastic process models (QPO) and damped random walk models (DRW) to explore the best methods by which to detect periodic, quasi-periodic, and stochastic variables in the OGLE survey. Because stellar flares are expected to contribute a significant number of ``contaminant'' transient events in large volume surveys like LSST \citep{kowalski2009}, determining how to identify and classify stellar flares in such surveys has a broad impact beyond the stellar flare community. Despite this, they are often not targeted for identification in the aforementioned alert systems.

Historically, studies of flares in single- to multi-object time series data have used simple automated routines to identify flares in these data as events in which 2 or more consecutive observations exceed the background by N-$\sigma$, and sometimes relied on subsequent visual inspection of these events to verify accuracy (e.g. \citealt{davenport2014, silverberg2016}). Larger volume surveys comprised of high quality, quasi-uniformly sampled space-based data have explored basic flare template cross-correlation matching to de-trended data \citep{davenport2016} and machine-learning techniques \citep{vida2018} to better automate flare detections in large datasets; however, \citet{vida2018} have noted some contamination issues with some of these approaches. \citet{kowalski2009} was able to identify flare events in sparsely sampled SDSS observations of Stripe-82 by taking advantage of the multi-color nature of these drift-scan data. The best method to identify flare events in current and future irregularly sampled, single (or multiple) filter ground-based all-sky surveys has not been addressed.

In this paper, we explore different methods for identifying flare events from irregularly sampled ground-based time domain photometric survey data, focusing on template optimization using differential evolution. The results of the implementation of such a technique are presented here, along with PyVAN: the publicly available software created to carry it out\footnote{See: \url{https://github.com/kdlawson/pyvan}}. We discuss the problems facing flare searches in irregularly sampled data (Section \ref{sec:background}), and investigate common sources of contamination for prior automated flare searches (Section \ref{sec:contamination}). We then describe the technique that our software uses to identify flaring stars while also avoiding the most common sources of false positives in flare studies (Section \ref{sec:method}). This technique is then tested and calibrated by application to light-curves of simulated PTF-quality and known identity, generated by reducing Kepler short-cadence light-curves to PTF-like precision and sampling (Section \ref{sec:sims}). The recovery rate of the flaring status of stars in data like PTF's is probed by application of the PyVAN software to the PTF data of prior well-studied flare samples (Section \ref{sec:ptf_application}). Following this, we apply this technique to a selection of northern hemisphere PTF targets observable by TESS, producing a sample of strong candidates for TESS short cadence follow-up (Section \ref{sec:tess}). Finally, we briefly discuss application of the software to data of other recent or upcoming surveys (Section \ref{sec:additional_apps}).

\section{Data} \label{sec:data} 
The primary data analyzed in this study was collected at the Palomar Observatory over a period of approximately seven years as part of the PTF survey \citep{law2009}. PTF observations are taken in either SDSS g'-band or Mould R-band, with the majority of data being collected at an exposure time of 60s. Additionally, PTF data was complemented with data from the Pan-STARRS1 \citep[PS1;][]{magnier2013} survey taken in the g, r, i, z, and y broadband filters to allow for color selection of the PTF targets.

\section{Target Selection} \label{sec:sample}

To search for flare candidates in PTF data, the Large Survey Database \citep[LSD;][]{juric2012} is used to acquire PTF photometry meeting certain criteria. Since many PTF targets are observed in only a single filter, LSD is also used to cross-match any potential targets with Pan-STARRS-1 (PS1) data to allow for color-color selection. For analysis of SDSS Stripe 82 flares from \citet{kowalski2009} and Kepler flares from \citet{davenport2016} , PS1 $g-r$ and $r-i$ colors for targets are transformed to SDSS $r-i$ and $i-z$ colors using quadratic color transformations provided in \citet{tonry2012}, which are then dereddened using the Bayestar-17 3-D dust map \citep{green2018} to attain a minimum extinction value for initial selection. Unitless extinction values from Bayestar-17 are converted to $g-r$ and $r-i$ corrections using coefficients provided in \citet{green2018}. Following this, we implement the color selection criteria for M-type stars utilized by \citet{kowalski2009}. For TESS follow-up candidates, PS1 photometry is instead utilized to select targets bright enough for observation with TESS, requiring $I_C < 13$ \citep{ricker2015}. For every sample, PTF observations in the database are rejected if flags indicate bad astrometry, bad photometry, or the presence of halos or ghost pixels in the exposure. To avoid selection of light-curves which lack statistically significant photometric enhancements, targets are required to have at least one observation that is 3+ standard deviations brighter than the mean magnitude (a pre-computed light-curve statistic that can be used for selection in LSD). 

Once light-curves are acquired, further cuts are made as they are searched to select flare event candidates. The median absolute deviation (hereafter $\sigma_{med}$) is computed for each target light-curve. This value is then used to select flare event candidates using a variant of the consecutive outlier test \citep{hawley2014}. Any instance of a flux enhancement that is at least 5$\sigma_{med}$ brighter than the median value with a neighboring point at least 2.5$\sigma_{med}$ brighter is recorded as a candidate flare event to use in fitting. Targets having at least one such candidate event remain in the target pool.

\subsection{Kepler Flaring Targets}
The sample of 4041 Kepler flare star candidates from \citet{davenport2016} is reduced to 158 targets by selecting for only the targets present in both PTF and PS1 data in LSD and which meet our color selection criteria for M-type stars. From these, our database query returns a sample of light-curves containing 1,877 R-band observations of 23 targets and 1,860 g-band observations of 26 targets, with 41 unique targets in total (117 are eliminated by our query cut requiring at least one bright outlier in the light-curve). Following application of the consecutive outlier candidacy test, 4 R-band and 6 g-band Kepler flare candidate light-curves remain for template fitting.

\subsection{SDSS S82 Flaring Targets}
Rather than starting with the 236 flaring M-dwarf stars in the \citet{kowalski2009} sample, the entirety of the Stripe 82 region in PTF is searched for M-type candidates. The full region is selected here as both PTF and SDSS data feature irregular sampling, making the recovery of flaring targets in PTF that were not identified in SDSS likely. The initial selection of Stripe 82 targets in PTF contains 4,194,553 R-band observations of 37,113 targets and 1,619,796 g-band observations of 18,686 targets, with a total of 54,482 unique targets overall. Following the consecutive outlier selection process, 2,804 R-band light-curves and 1,934 g-band light-curves for the Stripe 82 targets remain.

\subsection{TESS Follow-up Candidates}
For the sample of TESS follow-up candidates, LSD is used to select for any Northern hemisphere targets meeting the aforementioned criteria and observable with TESS. Following the initial query, the selection includes 6,354,757 R-band observations of 174,707 targets and 8,315,543 g-band observations of 157,682 targets. After eliminating targets which fail the consecutive outlier test, the final sample includes 3234 R-band light-curves and 2163 g-band light-curves.

\section{Identifying Flare Events in PTF Data}\label{sec:background}
Early in this investigation, attempts were made at identifying likely flare stars by use of pre-computed statistics for each PTF light-curve contained within LSD. In total, two-dimensional scatter plots comparing 10 statistics were visually inspected to search for promising separations of flaring stars from the rest of the stellar population and from other varieties of variable stars. The examined light-curve metrics are described in Table \ref{tab:metrics}. Ultimately, a number of statistics were identified which tended to differentiate some members of known variable star populations from the general population (Stetson J especially). However, none of the tested relationships reliably separated known flaring stars from other types of variable stars. A substantial roadblock for this approach was the difficulty of finding a sample of target light-curves in PTF in which verifiable flare events occur. While previously identified flaring stars can be selected in PTF, there are a number of confounding issues that remain: 

\begin{enumerate}
\item A star with flares observed in some other survey may not have coverage of any flare events in PTF's data and will then manifest more similarly to an inactive star in light-curve statistics

\item An apparent bright enhancement in the PTF data of a known flare star may still be produced erroneously by PTF's often large photometric errors or by issues with PTF's automated photometry. Without contemporaneous observations of the target star by another survey, these possibilities are not easily distinguishable

\item Blindly adopting prior samples risks propagation of contamination. Training new techniques with prior samples that may contain contamination can result in a new technique that is specifically engineered to accept the false positives of its predecessor
\end{enumerate}

These issues make the testing of a technique for flare finding in data like PTF's especially difficult and introduce a number of necessary design constraints for any such approach. Namely, the method must be able to recover flares while rejecting flare-like events being contributed both by photometric noise and common astrophysical contaminants (see Section \ref{sec:contamination}). Further, this should be achieved without training such a technique under the assumption that flare-like features in PTF light-curves of previously identified flaring stars are truly flares. 

\begin{deluxetable}{cp{12cm}}\label{tab:metrics}
\tablecaption{Light-curve metrics inspected for flare star separation in PTF}
\tablehead{
\colhead{Metric} & \colhead{Description}
}
\startdata
$\chi^2$ & Chi-square metric for observations in the light-curve \\
\
RMS & Root mean square deviation for the target's magnitudes \\
\
Con & Fraction of light-curve observations in which three consecutive observations are at least twice the RMS from the median magitude, plus 1 \\
\
Max Slope & Maximum light-curve pairwise slope across all consecutive observations \citep{richards2011} \\
\
Stetson J & A measure of the degree of autocorrelation among the light-curve magnitudes \citep{stetson1996} \\
\
Stetson K & A robust measure of the kurtosis for the distribution of magnitude measurements \citep{stetson1996} \\
\
Peak Significance & The difference between the median and minimum magnitudes, divided by the target's RMS \\
\
Amplitude & The difference between the target's maximum and minimum magnitudes \\
\
Modified m-statistic  & The difference between the median and minimum magnitudes, divided by the target's amplitude; Modified from \citet{kinemuchi2006} \\
\
$A_{95-5}$  & As amplitude, but using the $95^{th}$ and $5^{th}$ magnitude percentiles instead of max and min \\
\enddata
\end{deluxetable}

\section{Flare Catalog Contamination}\label{sec:contamination}
As a next step, we look to prior automated flare searches to gauge what types of astrophysical variables, if any, appear frequently as contaminants. To do this, we examined the sample of flare stars from the Kepler flare catalog \citep{davenport2016}, perhaps the largest catalog of flare stars currently available. First, we checked the sample for overlap with the list of eclipsing binaries from the Kepler eclipsing binary catalog \citep{kirk2016}, as well as overlap with a list of known RR Lyrae stars in Kepler \citep{nemec2013}. Additionally, we conducted a literature review to identify variable stars of other classifications present in Kepler, including: Delta Cepheid, W Virginis, Delta Scuti, Mira, RV Tauri, Beta Cepheid, Gamma Doradus, $\alpha^2$ CVn, and ellipsoidal variables \citep{debosscher2011}. We found no overlap between the flare sample and the identified $\alpha^2$ CVn, Mira, RV Tauri, and W Virginis variables. Each of the other groups had at least some overlapping membership with the Kepler Flare catalog. The largest contributors of potential contaminants were found to be eclipsing binaries (640 targets), Delta Scuti stars (284 targets), Gamma Doradus stars (135 targets), and RR Lyrae stars (40 targets). With 4041 targets listed in the Kepler flare catalog, these four types make up 27\% of the full sample. 

Of course, it is possible that the non-flaring identity is erroneous (rather than the flaring identity) or that some of these targets do truly exhibit flares. However, the appearance of nearly every RR Lyrae from the comparison list (all but the eponymous RR Lyrae itself) and the lack of mechanisms understood to produce flares in objects of this type suggest a systematic misidentification. Moreover, this process was intended primarily to identify the targets for which automated flare searches were most likely to return false positives and is not intended to be rigorous in terms of precise numbers of contaminants of each type. Toward the goal of identifying reliable samples of flare stars, consideration of these frequent sources of contamination is necessary in both the design of detection techniques and in the testing of those techniques.

\section{The PyVAN Software}\label{sec:method}
PyVAN (\textbf{Py}thon \textbf{V}ariable \textbf{A}ssessment with \textbf{N}on-linear Template Optimization) is a parallelized, user-friendly, and publicly available software package for Python, created to aid in identifying flare star candidates in light-curves having irregular or sparse sampling and sizable photometric errors. PyVAN uses the differential evolution technique \citep{storn1997} to optimize fits of empirical light-curve templates to the data. For each template's determined best fit, the likelihood that it would produce the observed data is computed and comparison metrics are generated. These metrics can then be used to separate flare star candidates from a sample population. The software is intended to aid in identification of flare stars by reducing large samples of light-curves down to a quantity that can be more easily verified by inspection. The core components of PyVAN carry out the following tasks, as well as the initial identification of flare event candidates described in Section \ref{sec:sample}.

\subsection{Template Optimization with Differential Evolution}
LMfit is a non-linear optimization and curve fitting package for Python that implements variables as parameter objects, and offers a large number of pre-built optimization routines \citep{newville2014}. Using LMfit's parameter objects allows for easily specifying and altering bounds for each of the varying template parameters, ensuring that any returned fits are not only a statistical best fit, but also physically reasonable.

While LMfit was built to expand on the Levenberg-Marquardt least-squares optimization algorithm from SciPy, early testing in this application revealed that least-squares would explore only a small portion of each allowed parameter space, becoming trapped in local optimization minima. Instead, fitting is carried out using differential evolution. Differential evolution is an evolutionary optimization algorithm which, compared to least-squares optimization, requires more function calls but better explores the global parameter space. The typical implementation of differential evolution works by randomly initializing a trial solution population and then `evolving' each generation by adding a weighted difference of two candidates to an additional candidate, and accepting that new candidate if it is a better solution \citep{storn1997}. Each member of each generation's trial solution population also has some chance of being `mutated' to a different region of the parameter space, allowing population members to explore parameter values that might otherwise remain untouched. In this application, differential evolution allows for the best fit parameters for each template to be identified relatively quickly, despite parameter spaces that are often quite large as a result of sparse data that offers only minimal constraints. Additionally, it eliminates the burden of determining reasonable parameter initializations, instead requiring only bounds for each. For each template's optimization, we weight the light-curve's observations with the inverse of the photometric uncertainty.

\subsubsection{Flare Template}
The utilized flare template was empirically derived from Kepler observations of over 6100 flares \citep{davenport2014} (see Figure \ref{fig:templates}). This shape is expected to fit reasonably well to the majority of flare events, and is typically referred to as the `FRED' model, for `Fast Rise, Exponential Decay'. For ease of application to PTF light-curves, which are acquired in the form of magnitudes rather than fluxes, the software takes parameters in terms of magnitudes and converts these to relative fluxes internally. A target having a single candidate flare event is optimized in LMfit over a set of four parameters: quiescent magnitude, peak amplitude, flare start time, and rise duration ($m_0$, dm, $t_0$, and dt respectively). In the adopted model, the ratio of rise duration to decay duration (and thus total duration) is fixed. Given that rise duration is likely never measurable with any meaningful precision in PTF data, this parameter is more accurately a proxy for the flare decay duration or total duration. For studies analyzing very high cadence flare photometry, in which rise duration can be measured independently of decay duration, the PyVAN software can be altered trivially to fit for these durations separately. For the purpose of analyzing PTF data, however, adopting the template as presented is sufficient. Boundaries for the utilized parameters are set both for logical limitations (i.e., no negative durations or amplitudes) and for approximate observed limitations (to ensure that determined flare fits have plausible parameters).

For targets having multiple candidate flare events, the approach is somewhat more complicated. The candidate events are first sorted by their apparent amplitudes (descending). Then, up to three flare events are fit simultaneously, optimizing for a single quiescent magnitude as well as a start time, rise duration, and peak amplitude for each. By fitting multiple candidate events together, a single event is prevented from unduly influencing the placement of the quiescent magnitude, which could result in a value that poorly explains the remaining events. While fitting all candidate events simultaneously might produce a better result by ensuring that the quiescent magnitude which best explains the entire set of candidate events is found, doing so results in massively increased computation time for targets with large $N_{flares}$ in exchange for a negligible improvement (if any) in the resulting solution. In testing, fitting up to three candidates together was found to produce models indistinguishable from those resulting from larger numbers of simultaneous events for a more manageable increase in computation time. The solution found here is then used as the initial model for any remaining flare candidates, which are successively fit using only three parameters (sans quiescent magnitude) and appended to the model. Throughout this process, any flare candidates not already fit or currently being fit are masked to prevent them from affecting the results of other fits. When completed, the solution has $3N_{flares} + 1$ parameters, where $N_{flares}$ is the total number of candidate flare events in the light-curve. The number of peaks fit simultaneously can be altered using a keyword argument in the software. In many cases, the need for this procedure might be circumvented by adopting some version of a clipped average as the quiescent magnitude, eliminating $m_0$ from the set of optimization parameters. However, when sparse sampling results in a light-curve with an especially well sampled flare and poorly sampled quiescence, the implemented procedure achieves better performance without compromising general efficacy.

\subsubsection{Quiet Template}
To test the likelihood that an observed light-curve was produced by photometric errors alone, a simple flat line is also fit to each candidate, seeking the $m_0$ that optimizes the fit. For this purpose, least-squares optimization is sufficient, as the optimization's sole minimum will be the global minimum. For light-curves of many observations, such as those of Kepler, one would expect that the quiescent magnitude found in flare fitting and the one found here would be very nearly identical. However, with light-curves like those of PTF, they can diverge considerably. This occurs most radically for light-curves in which a small number of points are dominated by a single large-amplitude flare event or when observations of a flare event exhibit much smaller photometric errors. 

\subsubsection{RR Lyrae Templates}\label{sec:rrlyr_templates}
RR Lyrae variables are periodically pulsating variable stars with particularly sharp light curve features having large amplitudes \citep{sesar2009}. For this reason, contamination from RR Lyrae stars is of particular concern for flare searches. While more continuous data may allow for these targets to be filtered by applying techniques such as period folding, the poor time coverage of data like PTF's makes such detections uncertain. Moreover, the light-curves of flaring stars often exhibit periodic variability as a result of starspots, and may therefore result in strong responses in periodograms (with periods similar to those of RR Lyrae or other common contaminating variables). Using differential evolution to fit sets of empirically derived RR Lyrae templates to each light-curve, the possibility that a flare star candidate is an RR Lyrae star instead can be evaluated.

We utilized 23 g-band and 22 r-band templates that described the majority of RR Lyrae analyzed in \citet{sesar2009}. Each filter set includes two templates for RRc targets, with the rest being for RRab targets. Though created from SDSS observations, the chosen templates are found to fit RR Lyrae in PTF sufficiently for our purposes. The templates are acquired in the form of time and magnitude coordinates covering one period of the RR Lyrae variation. These are implemented in PyVAN by interpolating over the provided values to create functions which can be passed a set of observation times and RR Lyrae parameters, and which return a corresponding set of magnitudes. The four parameters --- base magnitude, amplitude, phase offset, period ($m_0$, dm, $t_0$, and dt respectively) --- are again allowed to vary over ranges loosely constrained by observed RR Lyraes.

\begin{figure}
\gridline{\fig{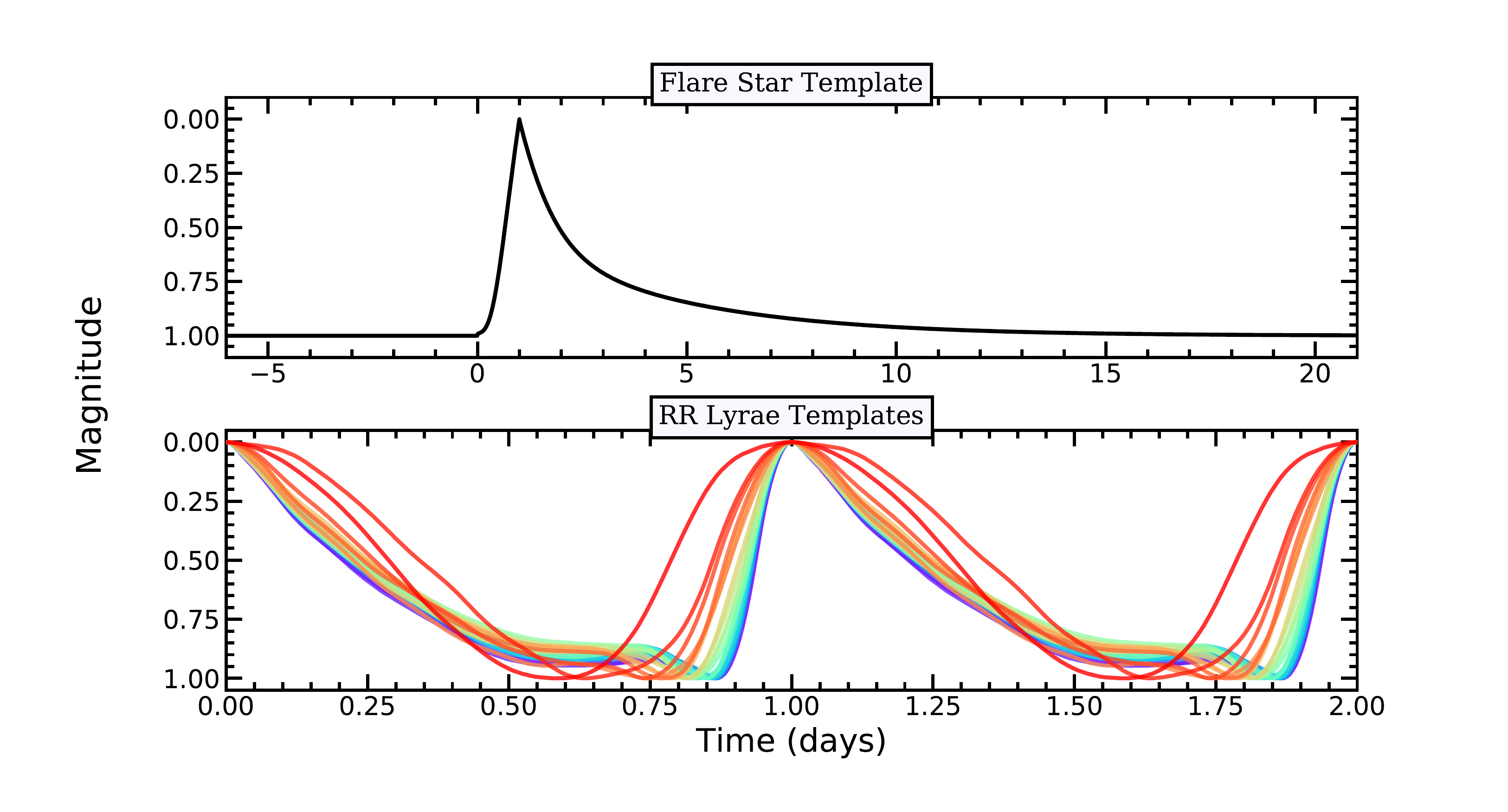}{0.95\textwidth}{}\label{fig:templates}}
\caption{\textbf{Top}: The empirically derived flare template from \citet{davenport2014} is parameterized in terms of quiescent magnitude, peak amplitude, flare start time, and rise duration ($m_0$, dm, $t_0$, and dt respectively). As depicted, the template has values of $m_0$ = 1, dm = 1, $t_0$ = 0, and dt = 1. \textbf{Bottom}: Two periods of empirically derived g-band RR Lyrae templates from \citet{sesar2009}. These are parameterized in terms of base magnitude, amplitude, phase offset, and period ($m_0$, dm, $t_0$, and dt respectively). The depicted templates have values of $m_0$ = 1, dm = 1, $t_0$ = 0, and dt = 1. Notably, both the flare template and RR Lyrae templates feature tall, sharp peaks. This, combined with sparsely sampled data, can result in light-curves whose identities are difficult to distinguish.}
\end{figure}

\begin{figure}
\gridline{\fig{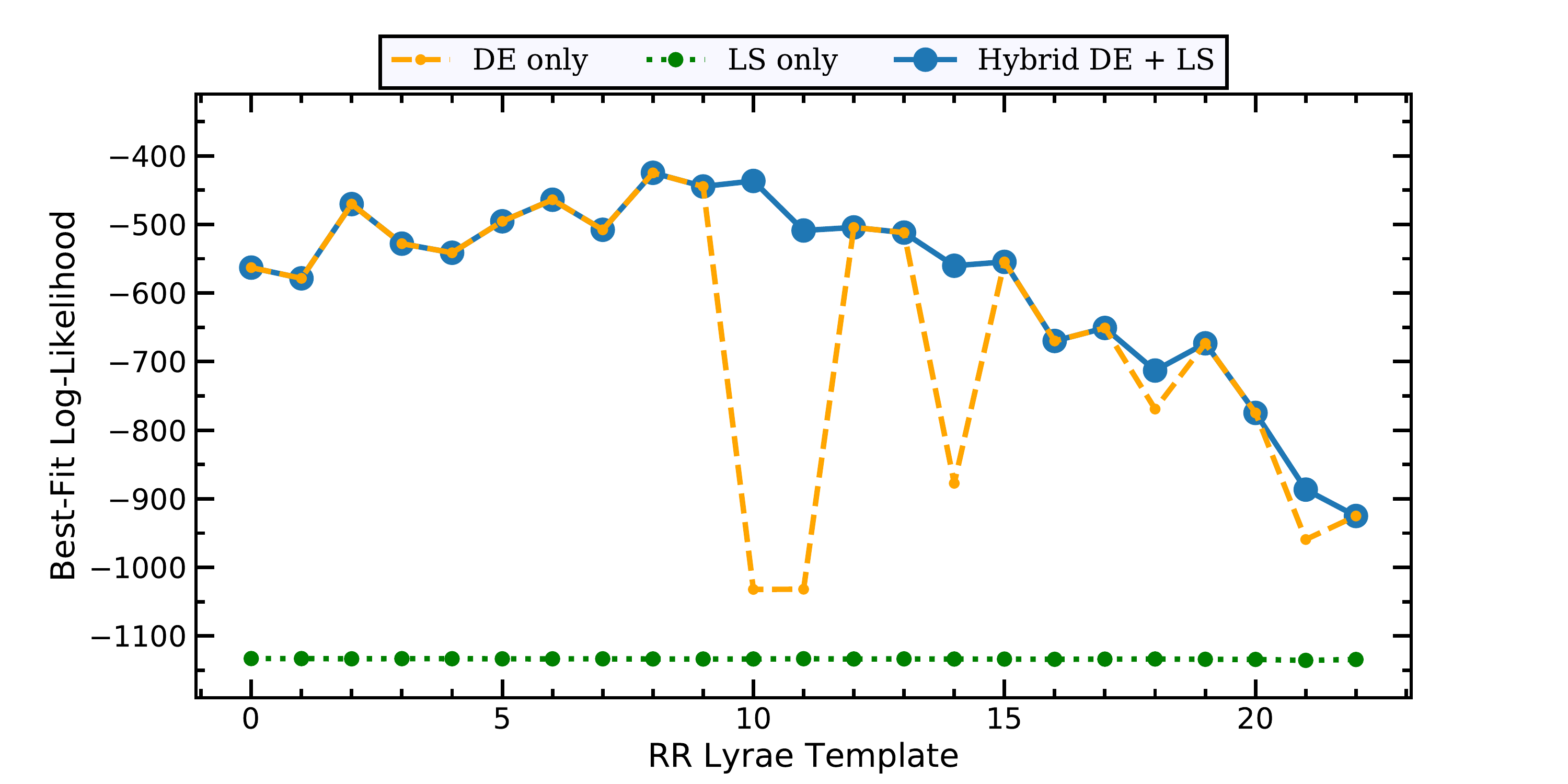}{0.95\textwidth}{}\label{fig:rrlyr_technique}}
\caption{A comparison of best-fit log-likelihoods for three different optimization techniques (see Section \ref{sec:rrlyr_templates}) used to fit 23 RR Lyrae templates to the light-curve of a known RR Lyrae star in PTF. Pure least-squares (green dotted line) provides a consistently poor fit, becoming trapped in local minima and exploring very little of the parameter space. Pure differential evolution (gold dashed line) generally provides good fits, but is time-consuming and sometimes converges to a sub-optimal solution (for at least 5 templates here). The hybrid technique (blue solid line) fits the first template repeatedly using DE and then initializes each subsequent template for optimization with least-squares using the best-fit parameters of its predecessor. This results in a more consistently strong fit and takes less time than a pure DE optimization.}
\end{figure}

For optimization, differential evolution alone proved to be time consuming and often resulted in poor fits, with results changing meaningfully from run to run and varying more between similar templates than expected (as a result of differential evolution's metaheuristic and stochastic nature). Instead, a hybrid optimization procedure is implemented, utilizing the fact that the shape of each subsequent template changes only slightly from its predecessor (see Figure \ref{fig:templates}). The first template is fit repeatedly using differential evolution (12 times by default), saving the results of the best fit from among them. Repeated fitting this way was found to produce superior results and with shorter computation times compared to altering the available optimization parameters. The first template's best fit is then used to initialize the second template's fit, which then uses least-squares optimization instead of differential evolution. The effectiveness of least-squares here depends on the fact that the global minimum in the parameter space will have moved only slightly for the new template, and that the local minimum that least-squares finds will actually be the global minimum for the new template. Following this, each template is initialized with the best fit parameters of the previous template, and optimized with least-squares. In testing, this process not only produces superior fits to the individual constituent routines (see Figure \ref{fig:rrlyr_technique}), but also requires much less time to employ than differential evolution alone. With typical photometric uncertainties, the final template choice may improve fit metrics only slightly (such that any one template can usually allow an RR Lyrae star to be distinguished). However, the processing time is dominated by the differential evolution portion such that fitting the first template takes many times longer than fitting all of the remaining templates using least-squares optimization.

\subsection{Comparison Metrics}
For each target, the best parameters for all three fit templates are stored, as well as a log-likelihood for each of these fits. We compute log-likelihood as a function of the $\chi^2$ of the residuals between the observations and the best-fit model as well as the number of observations in the light-curve:

$$
ln(L) \;=\; \ell \;=\; - \frac{n_{obs}}{2}\cdot ln\Big(\frac{\chi^2}{n_{obs}}\Big),  \:\:\:\:\:\:\:\:\:\: \chi^2 \;=\; \sum_{i=1}^{n_{obs}} \frac{(m_{obs, i}-m_{model, i})^2}{\sigma_i ^2}
$$

To identify strong flare candidates and eliminate likely false positives, the differences of best-fit log-likelihoods between each target's templates are computed. These include: (flare $-$ quiet), (flare $-$ RR Lyrae), and (RR Lyrae $-$ quiet), henceforth $\Delta\ell_{fq}$, $\Delta\ell_{fr}$, and $\Delta\ell_{rq}$ respectively. A strong flare candidate is expected to demonstrate meaningfully positive values in both $\Delta\ell_{fq}$ and $\Delta\ell_{fr}$. Though a true flaring star with large errors or unfortunate sampling may manifest with low values in these metrics, such a target is considered to be `unrecoverable' in that the observed target's identity cannot be confidently determined with the available information.

A candidate's $\Delta\ell_{fq}$ value is an indicator of the statistical significance of its flare event candidate(s); a large $\Delta\ell_{fq}$ indicates a small chance that the observed features could be produced purely by the recorded photometric error. Since the flare model consists of a quiescent model with one or more deviations to include outlier flare candidates, $\Delta\ell_{fq}$ should never manifest as significantly negative. 

Meanwhile, $\Delta\ell_{fr}$ serves primarily as an indicator of the ambiguity of the candidate light-curve's shape with respect to the adopted flare model. A significantly negative $\Delta\ell_{fr}$ indicates that the candidate's light-curve morphology is better explained by some aspect of an RR Lyrae template. A small value of $\Delta\ell_{fr}$ may indicate that the observed candidate event is perfectly consistent with the flare template, but that the information available is insufficient to rule out shapes consistent with the light-curve morphologies of RR Lyrae stars.

$\Delta\ell_{rq}$ is not utilized for flare identification. While it would likely be helpful in searching irregularly sampled data for RR Lyrae (in place of $\Delta\ell_{fq}$), three log-likelihood values allow for only two non-degenerate $\Delta\ell$ values (i.e., it provides no unique information for identifying flaring stars).

\subsection{Processing Times}
PyVAN carries out fitting a target for all desired templates on a single processor core, distributing candidate light-curves in a sample to multiple cores for fitting as available. A typical PTF light-curve with a single flare candidate is fit with a flare template and a quiet template in $\sim$0.6 seconds on a modern consumer processor. Meanwhile, fitting the set of RR Lyrae templates takes closer to 10 seconds per target. The long fitting time for RR Lyrae templates can be mitigated in practice by applying these only to targets demonstrating a promising $\Delta\ell_{fq}$ following flare and quiet fitting (as described in Section \ref{sec:ptf_application}). On average, over all of the samples discussed in Section \ref{sec:ptf_application}, this procedure allowed processing of around 0.3 targets per second per core for PTF light-curves.  This works out to around 21 hours on a 4-core system to process a large sample of $10^5$ PTF targets having candidate flare events.

\subsection{Software Features}
PyVAN is designed to be easily altered and appended, and delivers products as Python dictionaries. The software includes functions to carry out the core of the PyVAN data reduction process, including fitting of flare, quiescent, and RR Lyrae templates, as well as additional functions for plotting result metrics and displaying the template fits for individual targets.

In addition to carrying out the core features described above, PyVAN also implements a number of additional features, including many for visualizing fitting results. Users can easily generate scatter plots of computed $\Delta\ell$ metrics, as well as displaying an interactive light-curve for an individual target of interest (which can be selected based on a target's location in the scatter plots), with buttons to overlay any of the fit template solutions. PyVAN also includes functions to aid in the generation of test data (as in Section \ref{sec:sims}), in which a set of high-quality light-curves or light-curve models can be used to simulate data of similar quality to an input lower-quality data set. 

Further, PyVAN allows users to input additional templates to fit to data, allowing functionality for groups particularly concerned with other contaminants or interested in identifying targets other than flare stars. New templates can be implemented in one of two ways. The first is by using a generalized procedure to optimize a desired template, in which the user provides a set of parameters, parameter bounds, and a Python function for the template (which takes arguments of a time array and the template parameter values and returns a corresponding array of magnitudes). The second option allows a user to create a more nuanced template fitting procedure as a Python function (such as those implemented for our flare and RR Lyrae templates), which takes an array of data containing (at least) observation times, magnitudes, and errors, and returns a Python dictionary containing fit parameters and a log-likelihood value for the best fit to the data.

\section{Testing and Calibrating Using Data of Simulated PTF-quality}\label{sec:sims}
When analyzing poorly sampled data of a target of known identity there is no guarantee that any archetypal features will be discernible or real. To conduct a meaningful examination of PyVAN's techniques in application to data like PTF's, it is therefore helpful to simulate PTF quality in data for which the target's identity is known and where the underlying light-curve shape is much better constrained. This \textit{a priori} knowledge can then be combined with the metrics that result from our procedure to gauge the effectiveness of the approach and to inform selection of targets based on these metrics in later application to real data.

\subsection{Creating Data of Simulated PTF-Quality}
To begin this process, we utilize a selection of short cadence Kepler light-curves of targets indicated as flare star candidates in \citet{davenport2016}, including both true flare stars and a number of apparent contaminants (identified in Section \ref{sec:contamination}). These serve as the `donor' light-curves from which we will produce light-curves of PTF-quality. The pool of contaminants includes RR Lyrae, eclipsing binary, Delta Scuti, and Gamma Doradus stars, as these groups represent the most frequent contaminants to the Kepler catalog in terms of raw number. For each donor light-curve, a central flux value is computed by clipping any values further than 2 $\sigma_{med}$ from the median and computing the error weighted average of the remaining values. This central flux is then subtracted from each light-curve and a simple linear detrending process is performed to eliminate large systematic effects between quarters. All quarters of short cadence data are then combined into a single continuous light-curve and converted to relative magnitudes. Following this, each composite light-curve is split into separate light-curves anywhere that more than 5 days elapses without observations (typically where a full quarter or more of data is missing). Separating the light-curves this way results in approximately continuous data that can be more accurately convolved with PTF's sampling without eliminating observations. These light-curves are then placed into a pool corresponding to the target's suspected identity. Once each pool is created, we visually inspect every light-curve to verify the suspected identity, check that data from separate quarters are well aligned, and ensure that no true flare events appear in any of the contaminant pools.

Next, the initial sample of PTF observations described in Section \ref{sec:sample} (i.e., prior to elimination of targets not meeting our consecutive outlier requirement) is used to prepare pools of observing intervals, quiet magnitudes, and photometric uncertainties for simulating PTF quality data. For each light-curve, both the time between each successive observation and the $\sigma_{med}$ clipped quiet magnitude is calculated and stored. Magnitude uncertainties are prepared from the full set of observations by creating magnitude bins of width 0.02 at increments of 0.01 (such that they overlap) over the span of observations and storing the corresponding uncertainties. This allows a sub-pool of representative magnitude uncertainties to be attained for any given magnitude while also preserving the variance of PTF’s uncertainties. 

To simulate a PTF-quality candidate light-curve using the pools described above, we implement the following procedure. For the type of target being simulated, a prepared light-curve is drawn from the respective pool of donor light-curves from Kepler and the time of its first and last observation is noted ($t_0$ and $t_f$ respectively). Then, we draw a single quiet magnitude value and add it to the selected Kepler light-curve’s magnitudes. From the pool of PTF observation intervals, a random value is drawn, `dt', and the first simulated data point is placed at time $t_1 = t_0 + dt_1$. Successive values are drawn and added to the previous point’s time to create additional data points, ceasing at time $t_i$ when $t_i + dt_{i+1} > t_f$. Each simulated observation time is matched to an observation in the input Kepler light-curve, allowing the simulation observation times to be perturbed from their exact drawn position by 60 seconds in either direction to match with one of the donor observations, after which the observation times and magnitudes of the matched donor observations are stored. This allowed shift in time avoids more significant issues resulting from attempting to re-bin the Kepler data in exchange for a minor cost in terms of the accuracy of the simulation to true PTF data. Any drawn points falling more than 60 seconds from a real point in the input Kepler data are removed. Then, for every obtained magnitude value, an uncertainty (`$\sigma_{PTF}$') is drawn from the corresponding sub-pool of PTF uncertainties. Each simulated observation's final magnitude is drawn from a Gaussian distribution centered on the selected donor magnitude with standard deviation $\sigma_{PTF}$. While the input Kepler data is already displaced from its `true' position by its own photometric uncertainty, the reported uncertainties for utilized PTF photometry are typically more than 200 times larger than those of Kepler photometry. As such, the inaccuracy incurred by this approach is negligible.

For each desired light-curve, the process above, following the selection of an input Kepler light-curve, is repeated until the resulting light-curve has at least 10 observations and passes the consecutive outlier test described in Section \ref{sec:sample}. This helps to ensure that each simulated light-curve will be as comparable as possible to real data being fit by the software. The number of attempts required to produce each target is tracked throughout this process and stored for use in later analysis.  Hereafter, we refer to these Kepler-derived light-curves of simulated PTF-quality as simply "simulated light-curves" or "simulated data". In total, we produce a sample of 2500 simulated light-curves of each contaminant type, along with 5000 simulated light-curves of flaring stars.

\subsection{Comparison to Real PTF Data}

\begin{figure}
\gridline{\fig{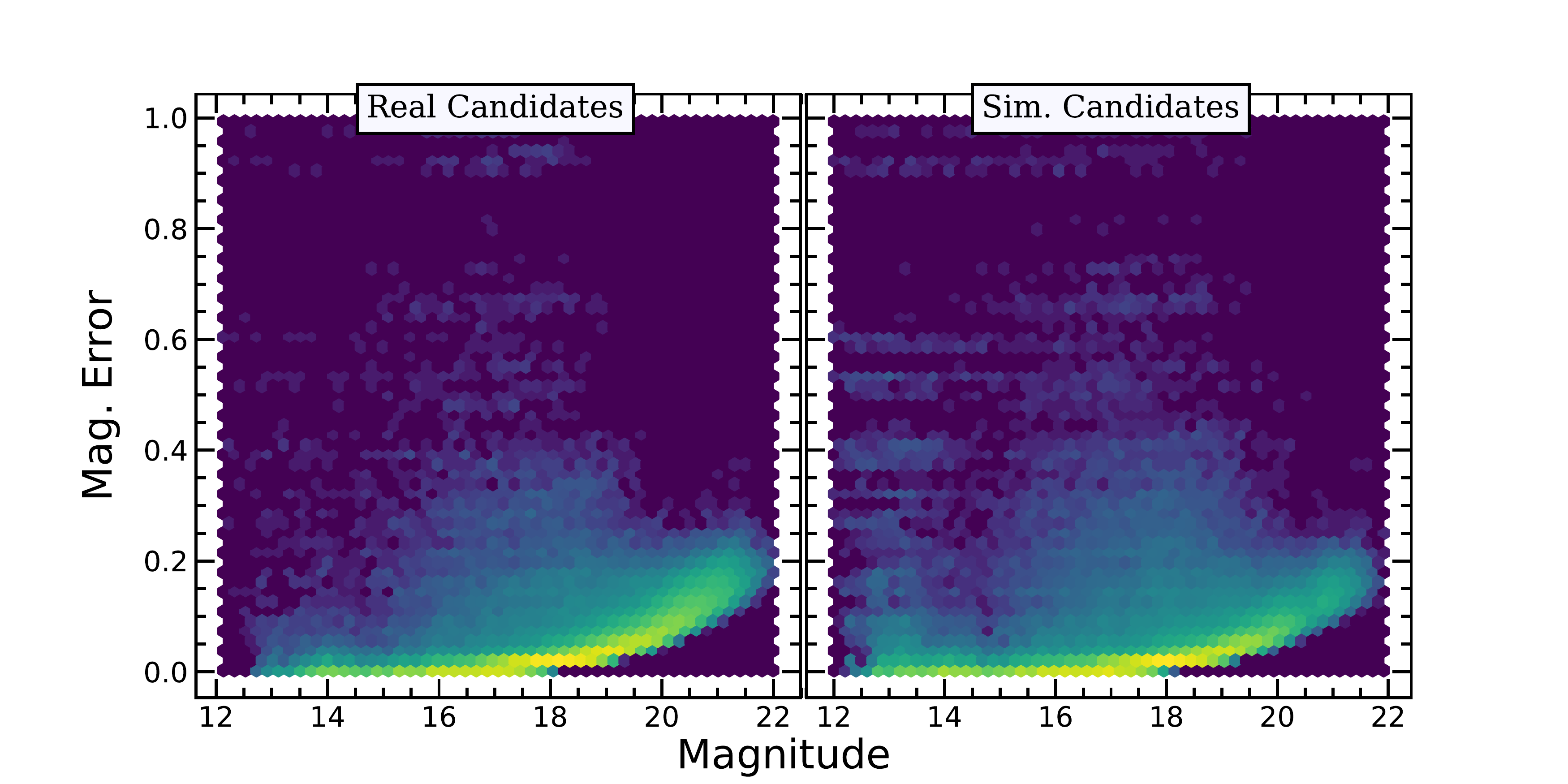}{0.9\textwidth}{}}
\gridline{\fig{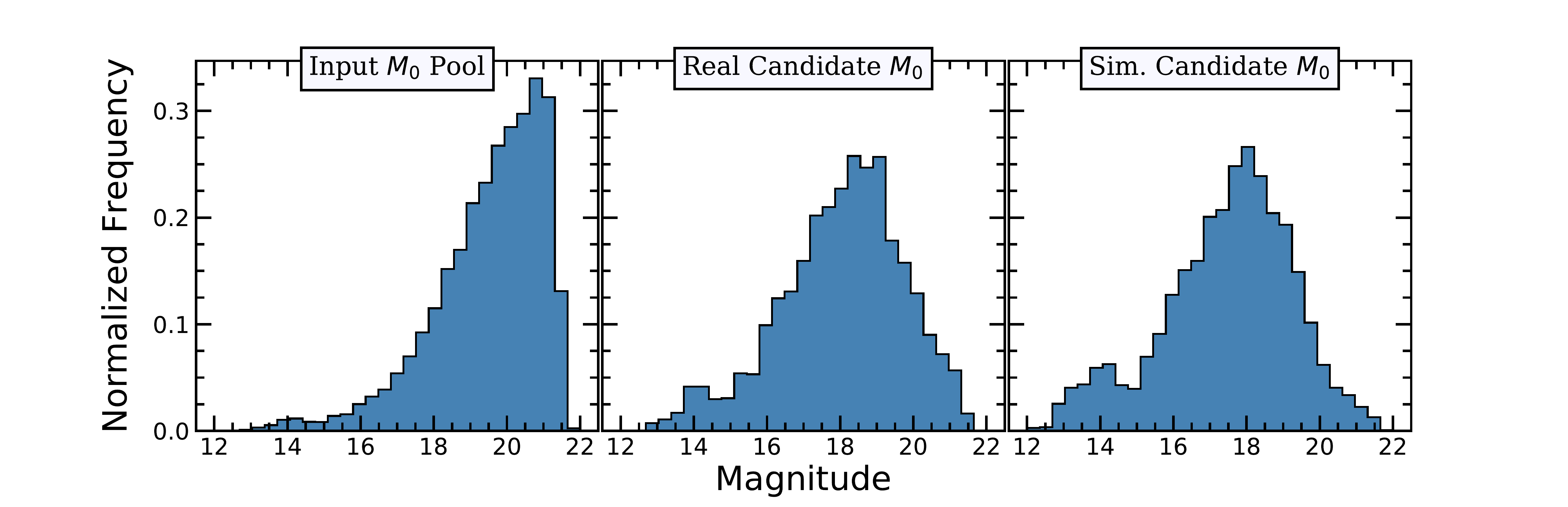}{0.9\textwidth}{}}
\caption{\textbf{Top}: Log-density plots of magnitude versus magnitude error for observations in both our sample of real flare candidates from PTF (left) and from our sample of simulated flare candidates (right) described in Section \ref{sec:sims}. \textbf{Bottom}: Histograms of: the pool of quiet magnitudes from the full set of queried PTF light-curves used in simulating data (left), the quiet magnitudes in our real sample of PTF flare candidates (middle), and the quiet magnitudes in our sample of simulated PTF flare candidates (right). In both sets of plots, our techniques for simulating PTF quality data from Kepler light-curves seems to produce data with similar distributions to our real candidate sample. 
\label{fig:sim_comparison}}
\end{figure}

The described process results in data that is generally comparable to real PTF data. The decision to re-select a quiet magnitude for each attempt results in distributions of quiet magnitudes and magnitude errors that appear to be organically biased toward the same types of light-curves that we find in our real sample following application of the consecutive outlier test (see Figure \ref{fig:sim_comparison}). The most notable divergence of the simulated data results from the difference between the durations of the Kepler and PTF surveys.  The utilized short cadence Kepler data spans a few hundred days at most for a given target, while PTF data frequently spans over a thousand days. It may seem reasonable to circumvent this issue by simply taking a given Kepler light-curve as representative of the target's behavior, and to instead duplicate Kepler light-curves end-to-end to accommodate PTF's longer observing span. However, every type of object we are interested in testing has some degree of periodic flux modulation. At a minimum, this approach would require precisely aligning each set of observations to avoid disrupting this periodicity, which would otherwise weaken the periodic RR Lyrae template fit and thus delegitimize the results. More troublesome still, long term variations in star spot modulation of flare stars or in the amplitudes and periods of RR Lyrae stars (the Blazhko effect \citep{blazhko1907}) would be very difficult to account for. As such, the resulting simulated light-curves must either fail in replicating the typical number of observations or the typical sampling rate of a PTF light-curve. We chose to focus on best reproducing PTF's sampling rate, as the alternate solution introduces the possibility of producing simulated data whose identity is easier to determine than the real data as a result of higher observation density. To mitigate the effect of this, donor light-curves were chosen from among the most observed Kepler targets of each identity.

Additionally, we have made no effort to eliminate any Kepler-specific systematic effects (save for eliminating the jump discontinuities between quarters by applying a linear flux offset), or to induce any PTF-specific systematic effects in our simulated data. Since the flags for both surveys are incomplete, the effects of this, if any exist, are difficult to estimate.

\subsection{Applying PyVAN to Simulated Data}\label{sec:simresults}
\begin{figure}
\gridline{\fig{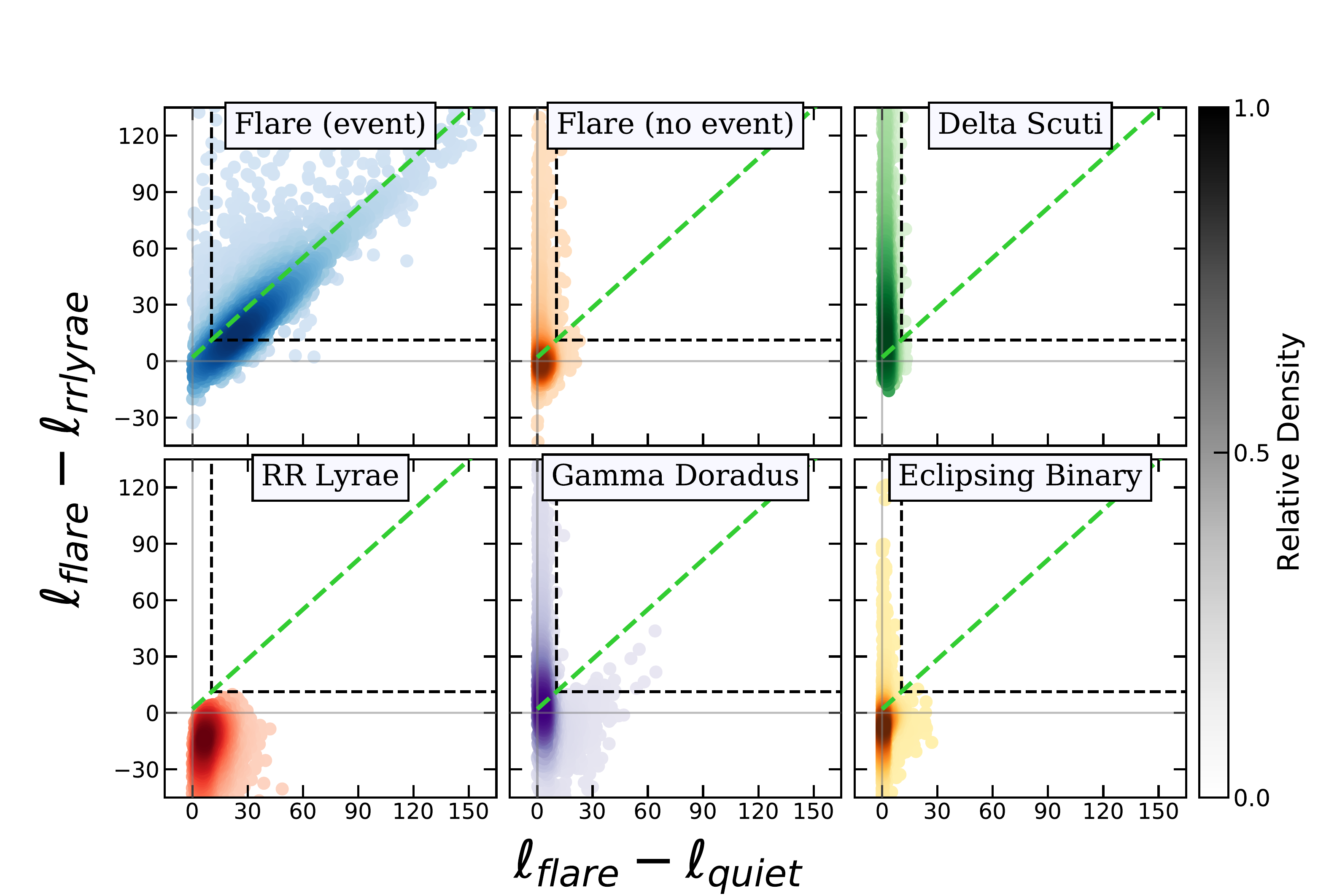}{0.58\textwidth}{}
		  \fig{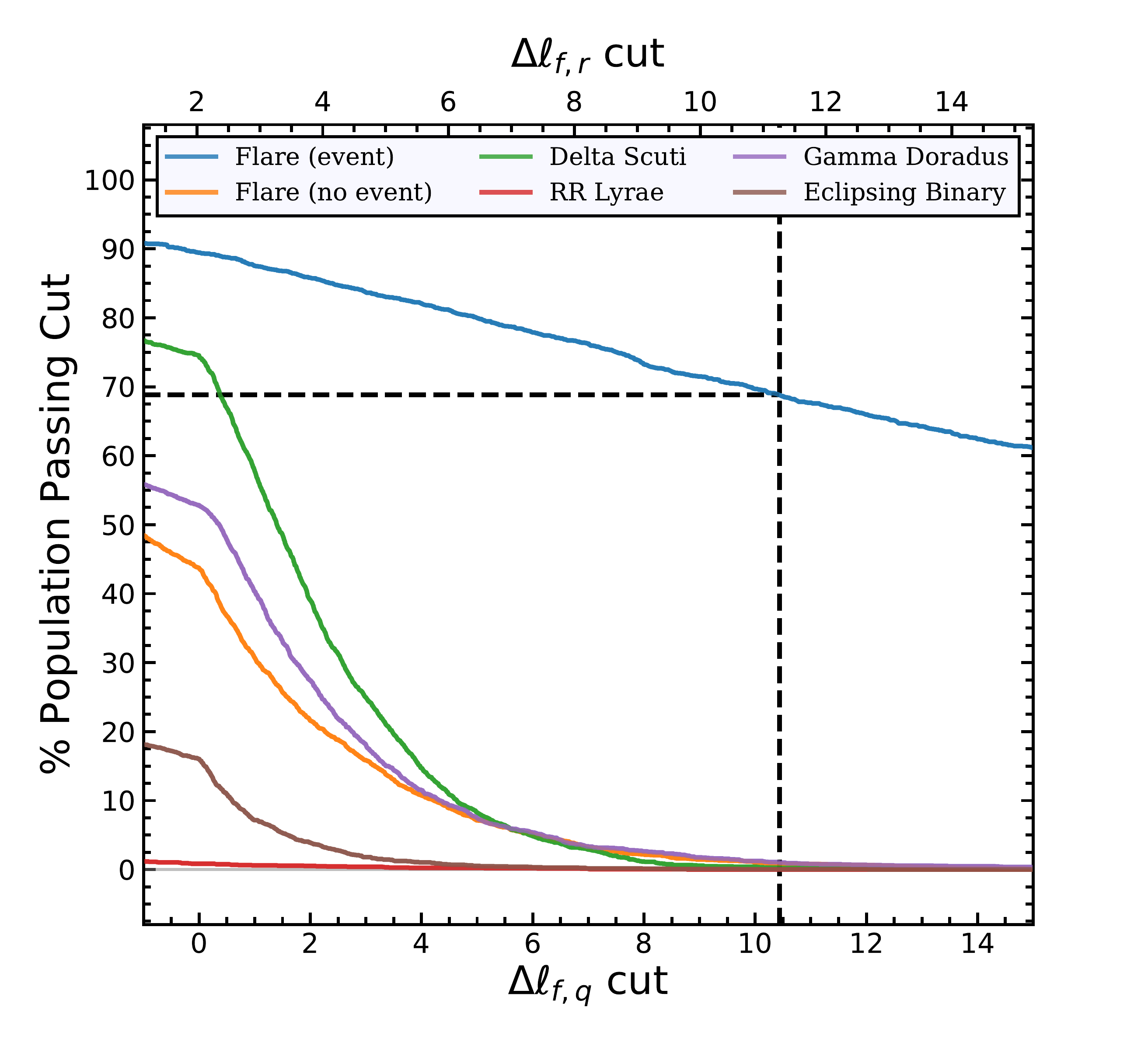}{0.40\textwidth}{}}
\caption{\textbf{Left}: Scatter plots of $\Delta\ell_{fq}$ versus $\Delta\ell_{fr}$ for fits to simulated PTF data, with color saturation corresponding to the relative density of the population (determined using kernel density estimation). The population of simulated flare stars is separated into two sub-populations based on whether or not their light-curves have coverage of a true flare event (``Flare (event)'' and ``Flare (no event)''). The dashed green line is the linear fit to the ``Flare (event)" population, while the dashed black lines represent the prescribed cuts on $\Delta\ell$ metrics above which \textless1\% of any contaminant population is found. \textbf{Right}: The percentage of each population included when applying a range of cuts extending from the dashed green fit line in the left plots. The $\Delta\ell_{fq}$ values of the cuts are along the lower x-axis, while the $\Delta\ell_{fr}$ values of the cuts are along the upper x-axis. The vertical dashed line corresponds to the \textless1\% contamination threshold depicted in the scatter plots to the left, while the intersecting horizontal dashed line gives the flaring population's throughput for that cut: 66\%.\label{fig:sim_scatter}}
\end{figure}

\begin{figure}
\gridline{\fig{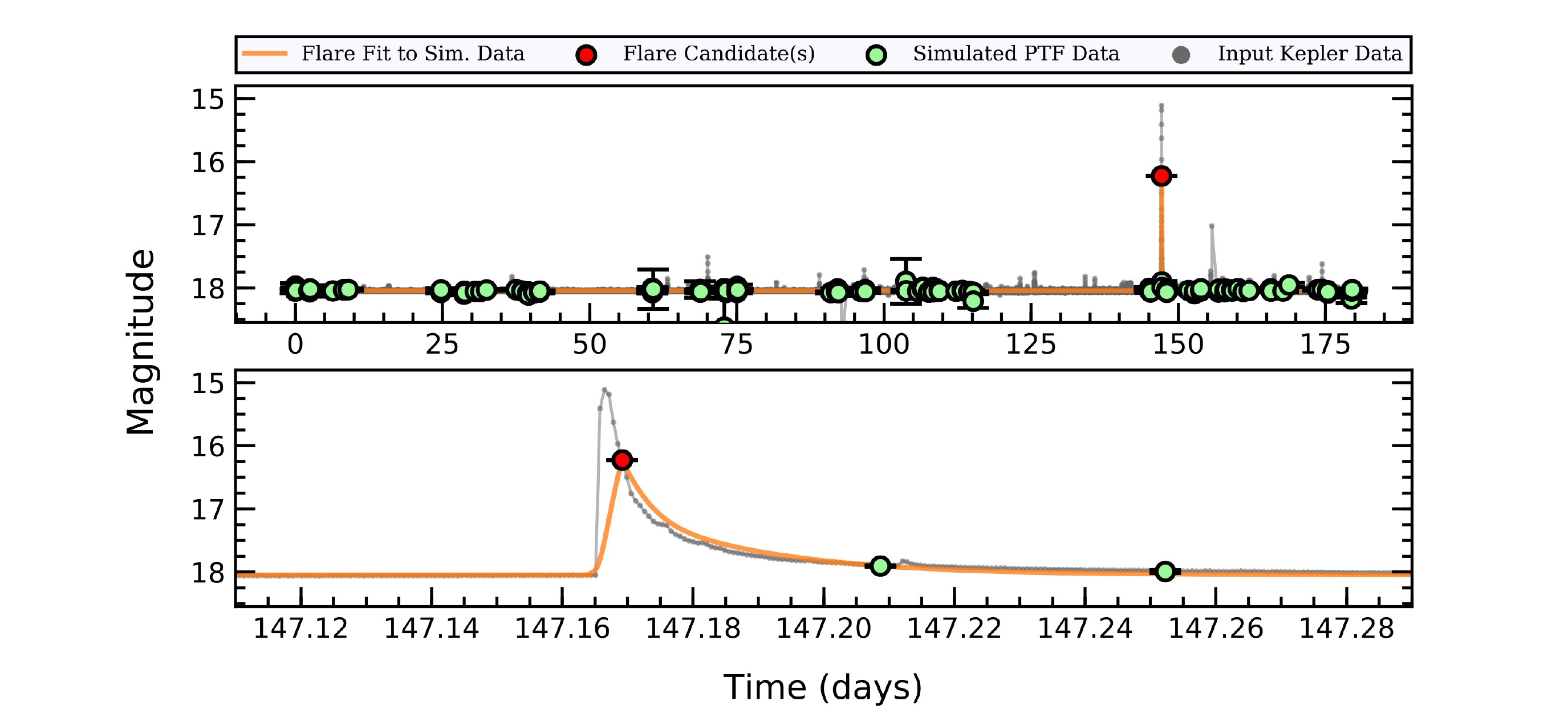}{0.95\textwidth}{}}
\caption{PyVAN flare fit to a simulated flare star light-curve (orange), where the lower plot simply features a magnified x-axis near the flare event. Gray points represent the donor Kepler data from which the simulated light-curve was created, while green points represent the simulated data, and with the identified flare event candidates shown in red. As a result of the uncharacteristically small uncertainties and the large amplitude of the event, fits to this target give large positive values for both $\Delta\ell$ metrics.\label{fig:sim_flare}}
\end{figure}

Applying the fitting algorithms and computing comparison metrics shows a meaningful correlation for simulated flare stars in the positive $\Delta\ell_{fq}$, positive $\Delta\ell_{fr}$ direction (Figure \ref{fig:sim_scatter}). An example of a strongly correlated flare fit to a simulated flare star can be seen in Figure \ref{fig:sim_flare}.

However, there exists a sizable population of flare stars whose fit metrics place them nearby the origin, mixing significantly with contaminant populations. Cuts can be made in this two-dimensional space to select a sample of targets containing flares with some inclusion of contaminant populations. Before deciding a selection criteria, we consider the different factors that are likely to result in weak metrics for a simulated flare star:

\begin{enumerate}
\item No sampled points coincide with a real flare event (sampling failure)
\item Sampled flare points have sufficiently large uncertainties so as to make recovery above a particular selection criteria impossible (quality failure)
\item Algorithm failed to identify a strong available model fit (optimization failure)
\item FRED flare model built onto linear quiescence poorly describes observed light-curve (model failure)
\end{enumerate}

We take steps as follows to delineate between these cases. This process is used to gauge the effectiveness of the software and to guide decisions as to what regions of the $\Delta\ell$ space should remain in the candidate pool following application to real data.

\subsubsection{Sampling Failure}
By design, simulating PTF data using much higher quality Kepler light-curves allows us to determine whether or not a real flare event was included in our simulated data. We define the locations of flares in the Kepler data as any point more than $3.5\sigma_{med}$ brighter than the median of the prepared donor light-curve. This criteria is an approximation intended to identify flare features which should be recoverable and will exclude any very small events or events embedded in star spot modulation, while also including some points which may not be feasibly recoverable with PTF's uncertainties. We then designate simulated flare targets as ``flare (event)'' or ``flare (no event)'' based on whether or not they include a detection of such an event. Henceforth, a flaring target refers to only those simulated light-curves containing coverage of an actual flare event.

The resultant sample of flaring light-curves includes 2101 targets, or 42\% of the original 5000 simulated flare star light-curves. At this point, we use linear regression to fit a line to the $\Delta\ell$ distribution for flaring stars. While outlier clipping could be used to identify a line that passes through the highest density region of the data, allowing the line to be displaced by the asymmetric distribution to the position depicted in Figure \ref{fig:sim_scatter} ultimately results in superior recovery rates. We then prescribe use of cuts in the positive $\Delta\ell_{fq}$ and $\Delta\ell_{fr}$ directions emanating from this best-fit line, where targets exceeding the threshold in both metrics would remain in the flare candidate pool (see Figure \ref{fig:sim_scatter}). As the position of this cut along the best fit line approaches the origin, both more flaring targets and contaminant targets will be included in the resulting sample. The ultimate position of the cuts should be determined by the number of candidates that the user is willing or able to visually inspect. For the purpose of subsequent application and analysis herein, we place these cuts such that 1\% or less of any single contaminant population is included. This results in cuts requiring $\Delta\ell_{fq}$ $>$ 10.44 and $\Delta\ell_{fr}$ $>$ 11.26. This region contains 69\% of the flare star simulations with an event, leaving 31\% of these (or 13\% of the full flare star sample) as unrecovered and so-far unexplained.

\subsubsection{Quality Failure}
As a result of PTF's often sizable photometric uncertainties, it is possible for a sampled flare event to manifest with uncertainties large enough to preclude the possibility of recovery above the chosen cuts by making the light-curve too statistically ambiguous for meaningful distinctions between templates. As with targets having no sampled flare events, we consider such a target to be unrecoverable using available information, and therefore an unmitigable loss. To identify cases where this has occurred, another metric is computed for the simulated flaring targets. In addition to the log-likelihoods for the flare, RR Lyrae, and quiet models, we add an `input' log-likelihood, in which the residuals between the original Kepler magnitudes and the simulation magnitudes, weighted by the simulation uncertainties, are used to compute the metric. This value represents the measurement that would be achieved if the software exactly recovered the input light-curve's original shape. The difference between the determined flare likelihood and this input likelihood, hereafter $\Delta\ell_{fi}$, serves as a measure of the quality of the determined flare model. A strongly negative $\Delta\ell_{fi}$ indicates that some model exists that would produce a much higher likelihood than the one that was found, while values near zero indicate that a good fit was achieved. A flaring target that was not recovered above our cuts is deemed unrecoverable if even with the input likelihood serving in place of the flare likelihood, the target still falls below the $\Delta\ell_{fq}$ and $\Delta\ell_{fr}$ thresholds. Applying this technique we find 336 such unrecoverable targets among the remaining flare star simulations, leaving 319 in one of the final two categories.

\subsubsection{Optimization Failure}
Still remaining are the possibilities that the PyVAN algorithm has failed to identify a strong available fit for the flare model, or that the utilized model has failed to adequately describe the flare event in the simulated light-curve. The former case would result if the differential evolution procedure converged to an inadequate solution because it failed to explore a preferable one or because it was restricted from a superior solution by our parameter boundaries. To test this possibility we open the parameter boundaries on flares to arbitrarily large or small values, increase the population of instantiated differential evolution test solutions dramatically, and then refit these targets repeatedly. Refitting each of these 319 targets 10 times results in 8 targets that improve sufficiently to pass our cuts. In each one of these cases the new solution is within the software's original parameter boundaries, indicating that the software simply converged to a sub-optimal solution initially. Comparing these light-curves with the set that passed without incident (1446 targets), we find that the number of observations in targets that were recovered only after refitting is $\sim$25\% lower on average. Based on this, the fraction of targets for which a poor solution is found may be inflated by the smaller number of observations typical in the simulated data. To minimize this effect in general application, simply refitting any targets falling near the thresholds would be sufficient. 

\subsubsection{Model Failure}
Though other factors may be to blame in isolated cases, the remaining 129 targets might be interpreted as failures of the utilized flare model to explain the observed behavior. Among other things, this can result from a flare event that has a distinctly different shape from the archetypal FRED model or from a quiescent profile that is poorly explained by a linear model due to starspot modulation (though the latter is rare given PTF's typical photometric uncertainties). In the first case, implementation of other flare model shapes could mitigate this loss, but may also serve to erroneously pass more contaminants. For instance, implementing a template for a Gaussian-shaped flare might also result in more frequently mistaking Gamma Doradus features (See Figure \ref{fig:contaminant_fits}) for flare events.

The latter might be mitigated by use of a detrending routine to eliminate periodic behavior before carrying out the flare fitting routine. However, without \textit{a priori} knowledge that the target is a flare star featuring significant spot modulation, it is difficult to determine when such a treatment should be applied or when it can be applied safely for PTF data. Implementing an automated detrending procedure for use on PTF data meeting some criteria is possible, but may well result in an increase of contaminants in excess of the small number of additional flaring stars recovered. Out of the 311 simulated flaring targets labeled as unrecovered due to model failure, Lomb-Scargle periodograms (\citealt{lomb1976, scargle1982}) identified 214 targets with periods having Lomb-Scargle powers over 0.25, the default threshold to warrant detrending used in \citet{davenport2016}. However, only 13 of these periods have a Lomb-Scargle power of over 0.25 in the original Kepler data, with all but 25 lying below even a power of 0.01. This suggests that the vast majority of strong periodicities identified this way will be inaccurate. While these ratios are likely affected by the reduced number of observations in the simulated data as compared to real PTF data, applying this test to only targets having 100 or more observations still results in only $\sim$20\% of detected powerful periods manifesting in the real data. Further, the longer time-span of observations for real PTF light-curves is more likely to be in excess of the lifetime of star spots, meaning that they cannot be assumed to manifest with stable periodicity. For general application, the failure of the chosen flare light-curve model for a small number of our simulated targets is accepted in lieu of any safe alternative.  

It should also be noted that some number of light-curves containing potentially recoverable flare events are also lost due to our candidacy requirement for at least two consecutive outlier observations. We exclude these targets to keep our resultant sample manageable and to avoid inclusion of candidate events that cannot be easily disentangled from non-flare causes, such as cosmic rays or errors in the automated photometry procedure.

\subsubsection{Flare Star Simulations Summary}
From 5000 simulated light-curves of flaring stars that pass our candidacy test, we recover flaring status in 1446 targets (29\% of the full sample) based on 1\% contamination thresholds. We find that 319 targets (6\% of the full sample) could have been recovered above the same thresholds with a different model or different model parameters, while 3235 targets (65\% of the full sample) are likely unrecoverable, due to the absence of sampled flare events or large uncertainties. If we instead consider only targets found to have recoverable flare events (1765 targets) we successfully recover flaring status for 82\%.

Combining this information with stored information regarding the numbers of simulated flare star light-curves that did not pass our initial cuts (and so were discarded), we can approximate the overall rate at which we expect to recover the identity of flaring stars in PTF data using PyVAN. We find that 5.5\% of all simulation attempts resulted in a target passing our cuts. Since we recover flaring status in 29\% of all simulations passing this cut, we can estimate the overall flare recovery rate as around 1.6\%. We note, however, that the lower number of observations typical in our simulated data reduces both of these measurements, and so the estimated recovery rate is probably lower than would be found in application to real data.

\subsubsection{RR Lyrae Simulations}
\begin{figure}
\gridline{\fig{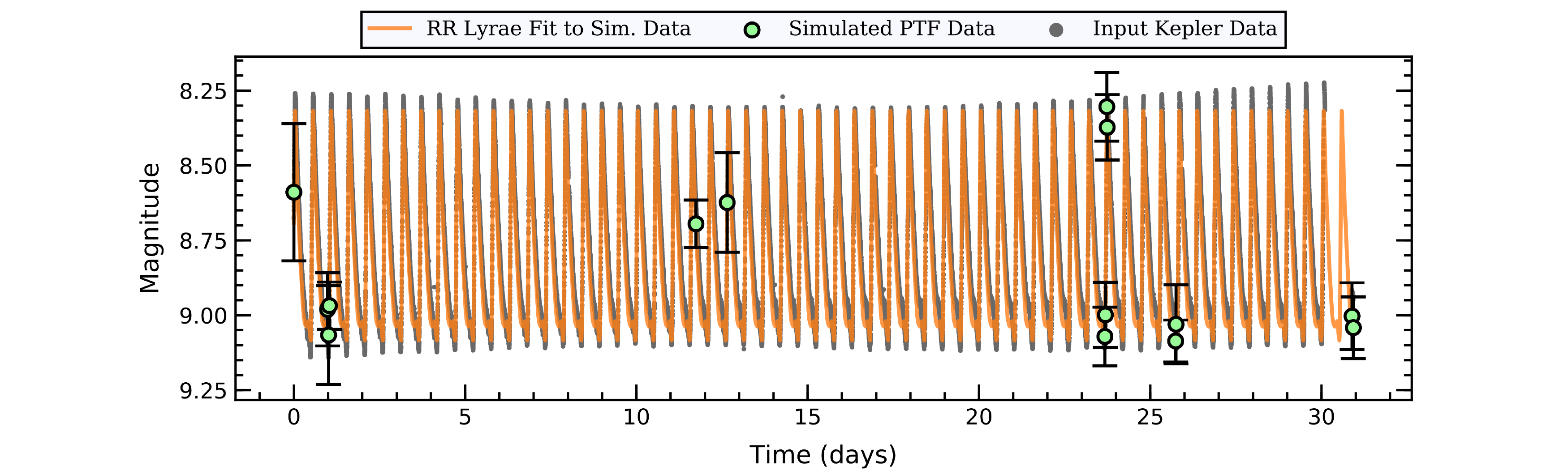}{0.9\textwidth}{}}
\gridline{\fig{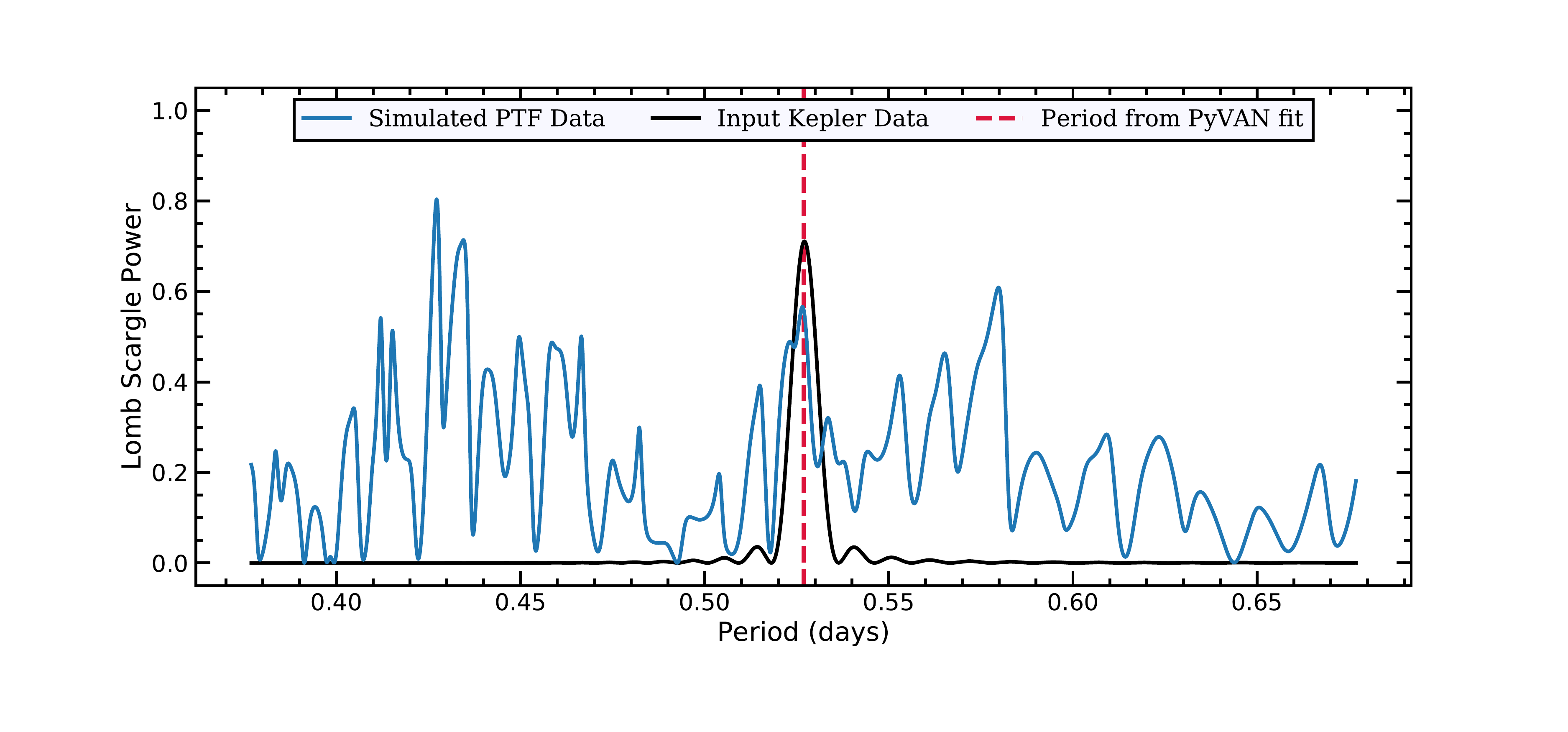}{0.9\textwidth}{}}
\caption{\textbf{Top}: PyVAN RR Lyrae template fit (orange line) to a simulated RR Lyrae target having only 14 observations over around 31 days. Gray points represent the donor Kepler data from which the simulated light-curve was generated, while green points represent the simulated data. \textbf{Bottom}: Lomb Scargle periodogram of the simulated data (blue) and the input Kepler data (black), as well as the period identified by PyVAN (red dashed line) for the simulated target above. Despite the inability of the Lomb-Scargle periodogram to detect the true period in the simulated data, the period determined by PyVAN template fitting falls within 19 seconds of the best Lomb-Scargle period for the full Kepler light-curve. By making a similar comparison for other RR Lyrae simulations, our testing indicates that the accuracy of the period from PyVAN's fitting constitutes a meaningful improvement over that of the Lomb-Scargle periodogram for light-curves having fewer than around 40-50 observations. This includes $\sim$20\% of targets in our samples of PTF targets overlapping Kepler and SDSS S82, and $\sim$70\% of targets for possible follow-up with TESS.}\label{fig:sim_rrlyrfit}
\end{figure}

The distribution of RR Lyrae in 2-D $\Delta\ell$ plots after fitting is meaningfully separated from that of the flaring population and contributes no targets as contaminants in the region above the \textless1\% contamination cuts. Despite the fact that the shapes of RR Lyrae in Kepler's bandpass are not perfectly described by the available RR Lyrae templates, the prescribed schema of $\Delta\ell$ cuts allows for recovery of more than 90\% of all simulated flaring light-curves before more than 1\% of the RR Lyrae simulations are included; the implemented procedure appears to effectively eliminate RR Lyrae stars as targets of concern in data of PTF quality. An example of a PyVAN RR Lyrae fit to simulated data can be seen in Figure \ref{fig:sim_rrlyrfit}.

\subsubsection{Results for Other Simulated Targets}
\begin{figure}
\gridline{\fig{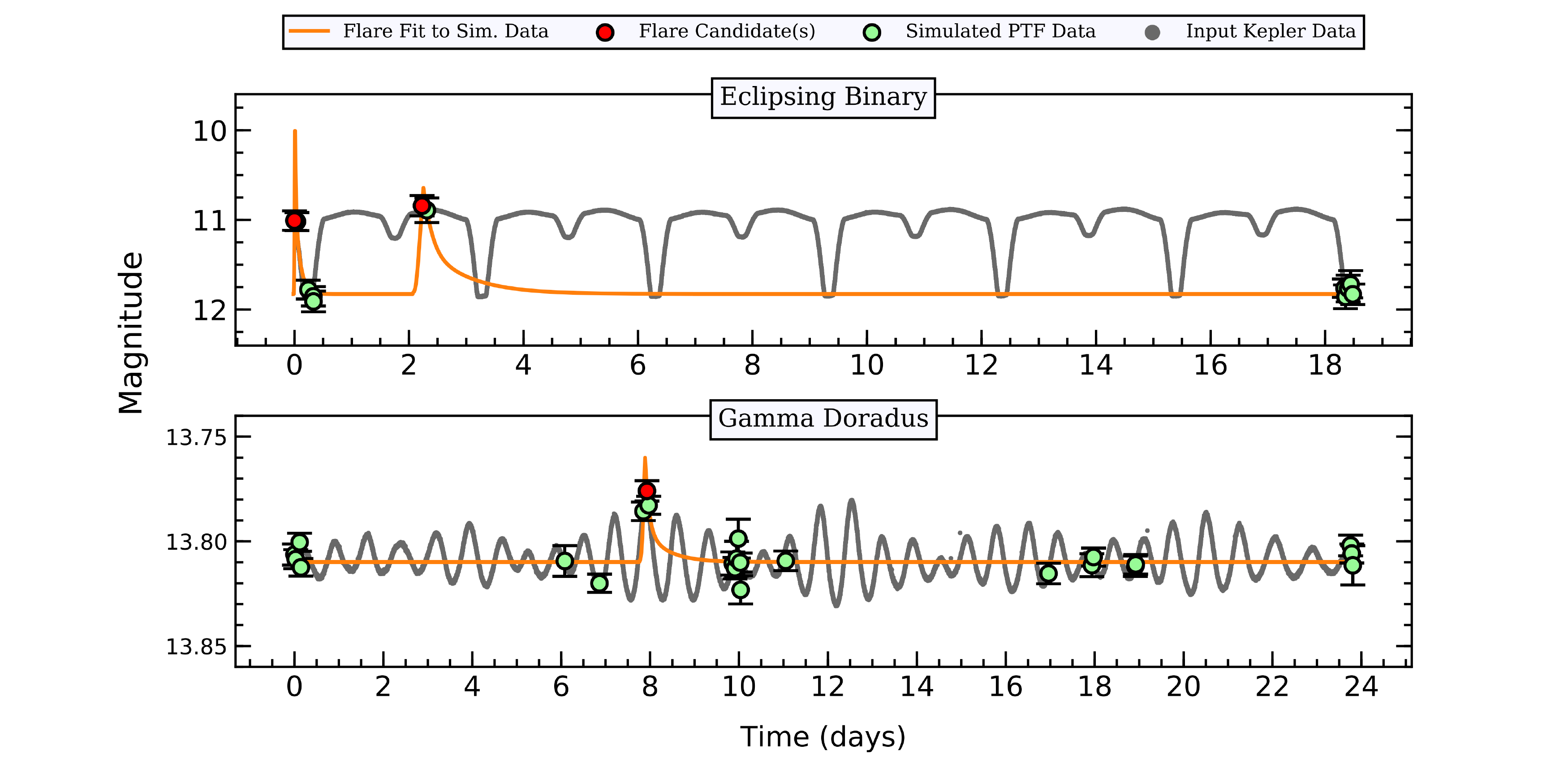}{0.95\textwidth}{}}
\caption{\textbf{Top}: Flare fit (orange line) to a simulated eclipsing binary light-curve. Gray points represent the donor Kepler data from which the simulated light-curve was generated, while green points represent the simulated data, and with the identified flare event candidates shown in red. As a result of the typically sparse sampling in PTF, the simulated light-curve manifests with flare-like features despite no such features being apparent in the donor light-curve. \textbf{Bottom}: As above, but for a simulated Gamma Doradus light-curve. By chance, the sparse sampling typical of PTF data results in mostly smaller amplitude features being sampled, with a single larger amplitude feature allowing the target to present with a flare-like light-curve. Additionally, the small overall amplitude of the flux modulation for the target precludes a good fit by RR Lyrae templates. Combined, these factors allow the target to pass into the regime of likely flare candidates.\label{fig:contaminant_fits}}
\end{figure}

Fits to simulated targets of other identities, including: eclipsing binaries, Delta Scuti, and Gamma Doradus stars tended to result in nearly Gaussian distributions in $\Delta\ell$-space, with centers lying slightly positive along $\Delta\ell_{fq}$ and slightly negative along $\Delta\ell_{fr}$. A small number of each population appears in the region dominated by flaring stars, with passing $\Delta\ell$ metrics. Typically, this seems to occur because of unfortunate sampling or large uncertainties that obscure their periodic nature. Examples of two passing targets from these populations can be seen in Figure \ref{fig:contaminant_fits}.

\section{Application to PTF data}\label{sec:ptf_application}
Template fitting techniques were applied to the selections of PTF flare candidates described in Section \ref{sec:sample}. To reduce processing times, each target was fit first for quiet and flare templates, allowing $\Delta\ell_{fq}$ to be calculated prior to RR Lyrae fitting. To remain in the pool of flare candidates, a target must have both a strong $\Delta\ell_{fq}$ and $\Delta\ell_{fr}$. As such, only targets having a $\Delta\ell_{fq}$ metric exceeding the identified $1\%$ contamination threshold of 10.13 (See Section \ref{sec:simresults}) are fit with RR Lyrae templates.

After fitting, we identified a recurrent issue with the data in which an apparent error in the zero-point calculation of PTF's automated photometry routine would induce a large amplitude flux enhancement for a small number of consecutive processed frames. This contributed flare-like enhancements to the light-curves of any targets in those processed images, and would occasionally allow targets to manifest with passing fit metrics; even though the enhancements may not be fit especially well by a flare template, a massive amplitude event in an otherwise quiet light-curve is likely to result in comparatively much worse RR Lyrae and quiet fits. To remedy this issue, in the full set of queried light-curves we counted the number of times that each processed image resulted in a 2.5$\sigma_{med}$+ outlier for a light-curve. We then flag any processed image resulting in an outlier event for more than 10\% of light-curves in which it appears, and resulting in at least two outlier events overall (to avoid eliminating single outliers in processed images that appear only a few times in the sample's light-curves). For each target light-curve containing a flagged observation, the observation is masked and the target is refit. Targets lacking a flare candidate following this process are eliminated from the candidate pool.

Further, we note the presence of non-astrophysical contamination that cannot be easily distinguished by light-curve analysis alone. Most commonly, this resulted from the presence of detector "ghosts". As these bright artifacts move across the detector, they induce time dependent flux enhancements for sources that they overlap. In some cases, these enhancements result in light-curve features that are fit very well by our flare template. The true cause of the apparent event is then only verifiable upon inspection of the source images. As such, we carry out manual inspection of source images to verify flare events from targets for TESS  follow-up or from the candidates overlapping with the Kepler and SDSS S82 flare samples.

\subsection{PTF Stripe-82 Fit Results}
\begin{figure}
\gridline{\fig{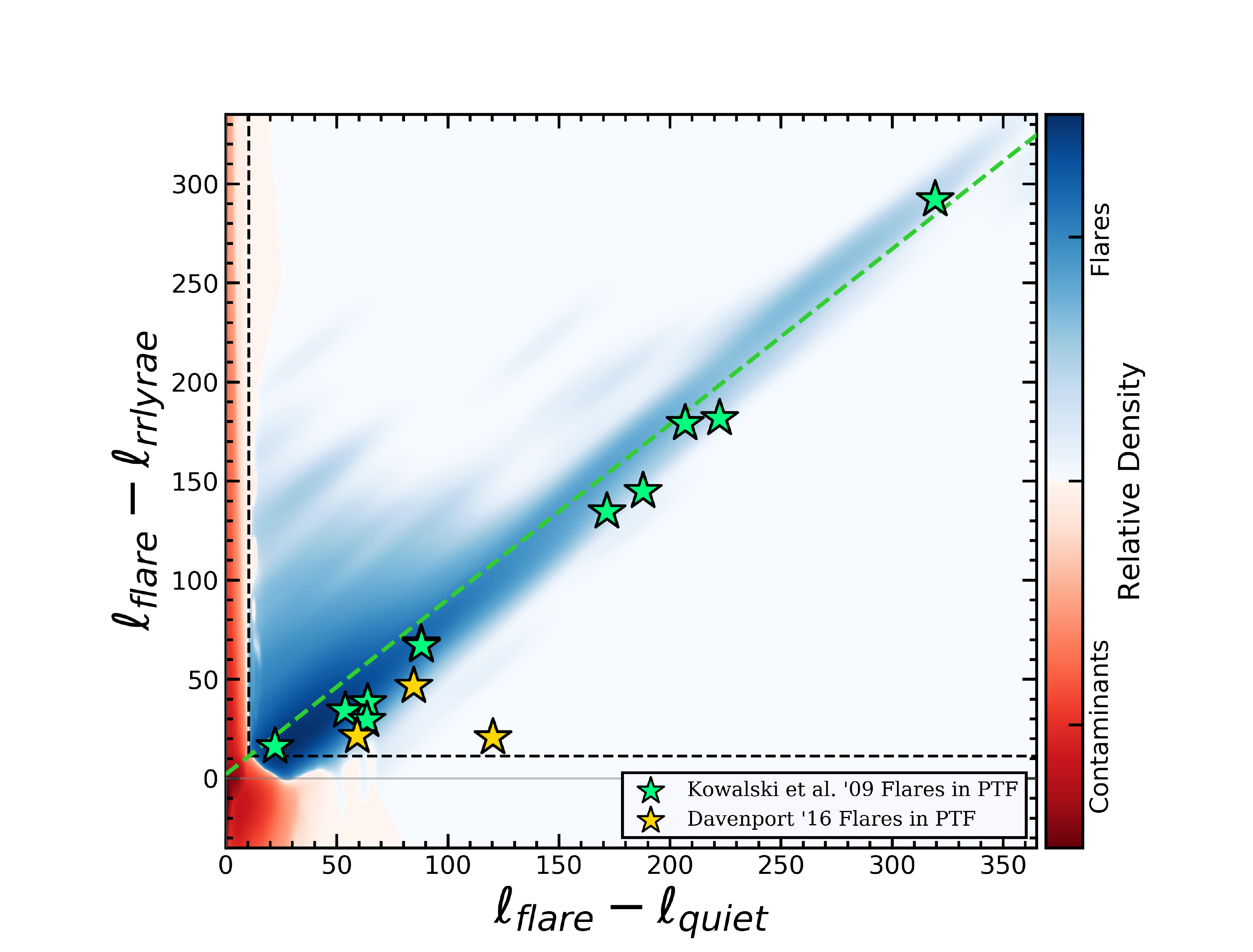}{0.495\textwidth}{}
		  \fig{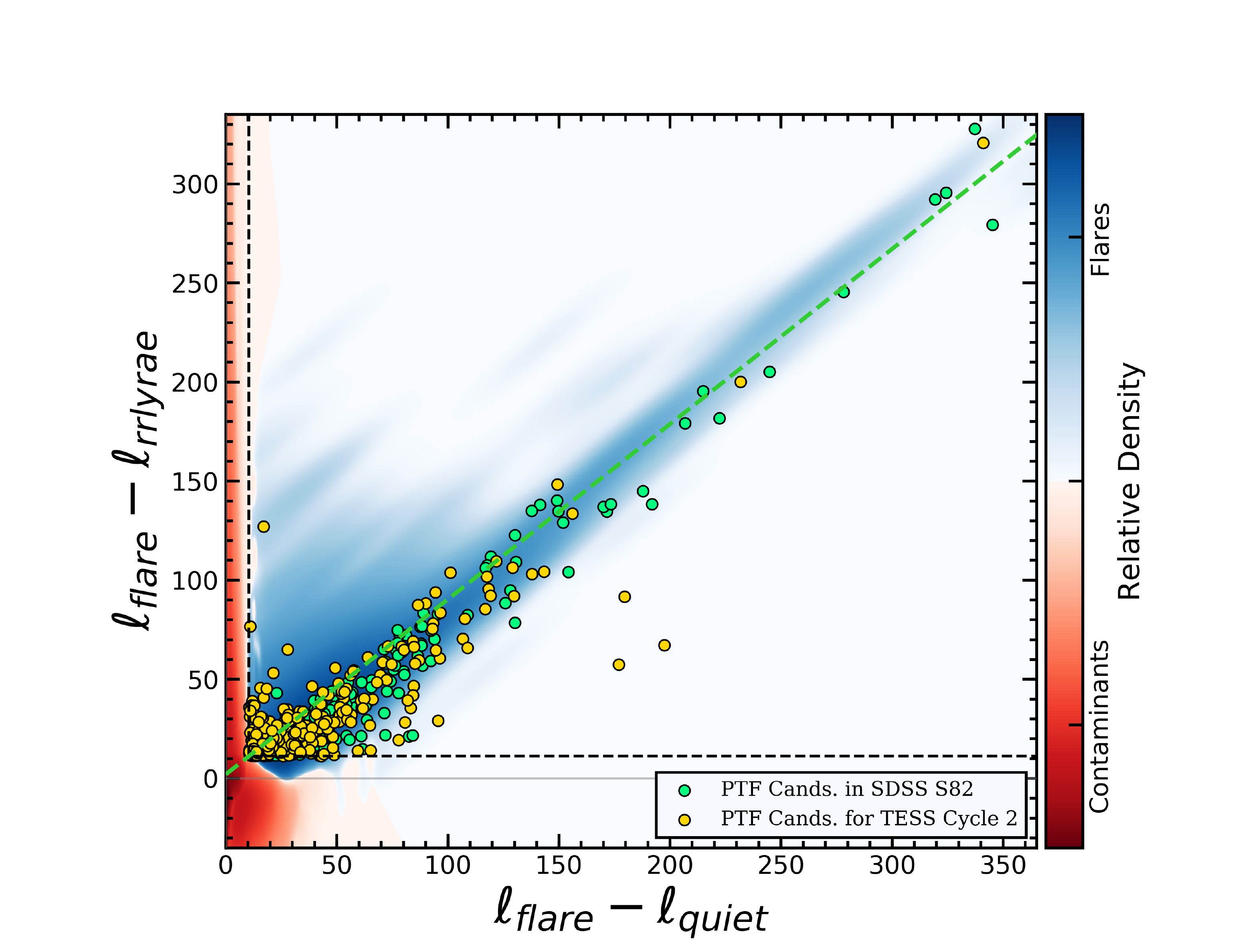}{0.495\textwidth}{}}
\caption{\textbf{Left}: $\Delta\ell$ values for fit PTF light-curves of recovered flaring targets from \citet{davenport2016} (gold stars) and \citet{kowalski2009} (green stars). Dashed black lines correspond to 1\% contamination thresholds determined in Section \ref{sec:simresults} using simulated PTF data, while the dashed green line corresponds to the best linear fit to the simulated flaring population. The background color indicates the relative density of simulated flaring stars versus the collection of all other simulated objects at that position (with bluer regions indicating more flaring stars than contaminants, and redder regions indicating more contaminants than flaring stars). Three targets from the Kepler flare sample and ten targets from the SDSS S82 flare sample are recovered (with a single SDSS S82 target being recovered in both g and R bands). \textbf{Right}: As left, but showing $\Delta\ell$ values for all flaring candidates with values exceeding our thresholds from the sample of M-dwarfs in SDSS S82 (green) and from the sample of targets considered for follow-up with TESS (gold).\label{fig:recovered_tars}}
\end{figure}

Of the 236 flaring stars in the \citet{kowalski2009} Stripe 82 sample, we recover 10 targets (4.2\%) above the thresholds determined in Section \ref{sec:simresults} using PTF's data, with a single target being recovered in both g and R bands (see Figure \ref{fig:recovered_tars}). Further, we identify an additional 245 flare star candidates for this region for which flare events were not identified in the prior SDSS S82 sample. Considering the similar recovery hurdles between the PTF data and SDSS data, the fact that our approach recovered a similar number of targets overall (236 versus 255, with 10 targets overlapping) is reassuring evidence that the technique is viable for future application to larger populations.

\subsubsection{Flare Luminosities}\label{sec:lums}
For an additional comparison with the results of \citet{kowalski2009}, we estimate peak flare luminosities for each target recovered from the sample of M-dwarfs in the SDSS S82 region. When the shape of the flare candidate is not perfectly constrained, the returned best-fit flaring log-likelihood is degenerate in that there are a range of fit parameters resulting in a model with the same likelihood. To compute reasonable luminosity estimates for identified flare events, we select the lowest amplitude flare fit which gives the approximate likelihood originally identified by the algorithm (requiring that the $\chi^2$ value of the fit using the new amplitude be within 1\% of the original fit's). In many cases, this constrains the amplitude to the brightest point observed for the event. However, when data includes rise phase observations, or exclusively slow decay phase observations, this `minimum' amplitude may be larger.

Luminosities are then computed following the general procedure of \citet{kowalski2009}. For each flare event, the target's extinction corrected quiescent flux is subtracted from that of the flare peak. Distances are determined using parallaxes obtained from the Gaia DR2 catalog \citep{gaiadr2,gaia}. Any target that isn't matched or that has an associated parallax uncertainty larger than 20\% of the measurement is eliminated from the candidate pool for the purpose of evaluating luminosity. The flare flux and distance measures are then converted to a g-band or R-band flare luminosity.

To better compare the luminosities of identified flare event candidates with those of prior studies, we compute flare luminosity conversion factors from PTF's g and R filters to SDSS's u filter. For this purpose, we follow the results of \citet{kowalski2018}, which concluded that the $\sim9000K$ blackbody approximation of flare continuum radiation from broadband photometry \citep{hawley2003} misrepresented the true spectrum of flares in many wavelength regimes. Instead, we compute the fluxes in each filter using the HST-1 flare spectrum provided with \citet{kowalski2018}. The ratio of the flux throughput in any two filters can then be used as an approximate conversion factor for a flare's luminosity between them. Multiplying a computed PTF filter luminosity by the ratio of this u-band flux and the respective PTF filter flux provides an approximate corresponding u-band luminosity. The determined conversion factors for PTF g-band and PTF R-band to SDSS u-band are 1.125 and 1.623 respectively.

\begin{figure}
\gridline{\fig{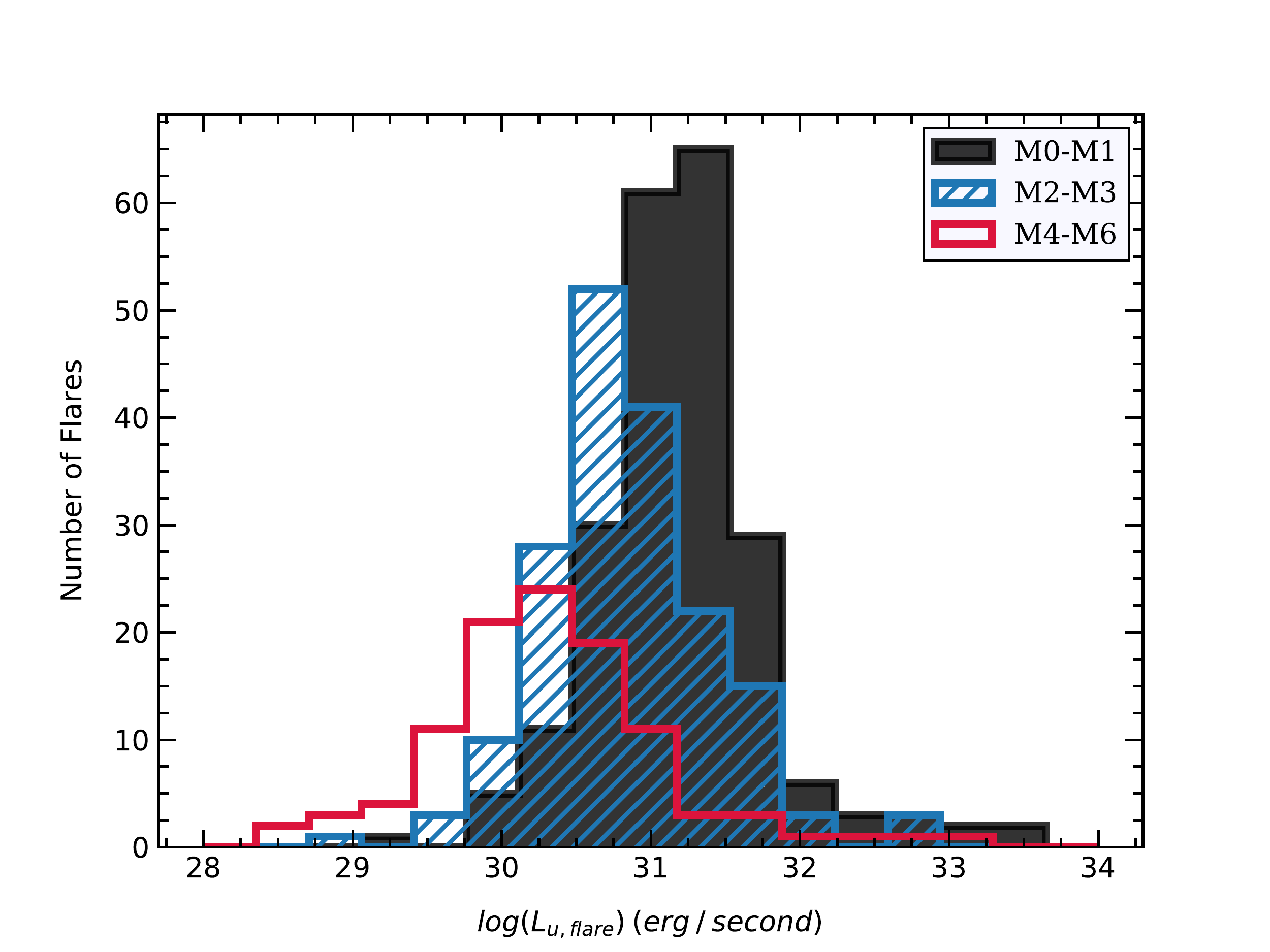}{0.9\textwidth}{}}
\caption{Flare luminosities for strong M-dwarf flare star candidates in PTF data of SDSS Stripe 82, identified using PyVAN. Luminosities have been converted from their native PTF filter to SDSS u-band using the ratios of filter convolutions with a flare spectrum from \citet{kowalski2018}. Targets are divided into three spectral type bins for comparison to prior results from \citet{kowalski2009}. Notably, our distributions appear to peak at higher luminosity than those of \citet{kowalski2009}. Since PTF's g-band and R-band have poorer flare-to-quiescent flux contrast than SDSS's u-band, a higher typical flare luminosity for recovered flare events is in line with expectation. \label{fig:luminosities}}
\end{figure}

After converting each PTF luminosity using these determined scaling factors, the two filters' samples are combined to produce a single set of u-band flare luminosities. For comparison with the SDSS flare sample, the set of luminosities is separated into three spectral type bins: M0-M1, M2-M3, and M4-M6. The resulting histograms (Figure \ref{fig:luminosities}) show luminosity distributions which peak at somewhat higher luminosities than those of \citet{kowalski2009}. However, as we use a similar candidacy test, but in bandpasses in which the contrast of flare flux to stellar flux is expected to be much poorer (especially so for PTF’s R-band), a higher average recovered flare luminosity is consistent with expectation.

\subsection{PTF Kepler Flare Catalog Fit Results}
Of the 158 M-type stars identified in the \citet{davenport2016} Kepler Flare Catalog, 3 targets (1.9\%) are recovered above thresholds determined in Section \ref{sec:simresults} (see Figure \ref{fig:recovered_tars}). Since Kepler’s photometry is generally of significantly higher quality than that of PTF or SDSS, the lower recovery rate for the targets in the Kepler flaring sample compared to SDSS S82 may be explained by targets exhibiting flares at a level recoverable in Kepler but that was not recoverable with PTF data.

Combined with the results of the SDSS S82 targets overlapping with the \citet{kowalski2009} sample, this corresponds to recovery of flaring status in $\sim3\%$ of tested flaring stars. Though slightly higher than our estimated recovery rate following Section \ref{sec:simresults}, this result is consistent with the expectation of a reduced recovery rate in our simulations due to the smaller average number of observations in our simulated data.

\subsection{PTF TESS Follow-up Candidate Fit Results}\label{sec:tess}

\begin{figure}
\gridline{\fig{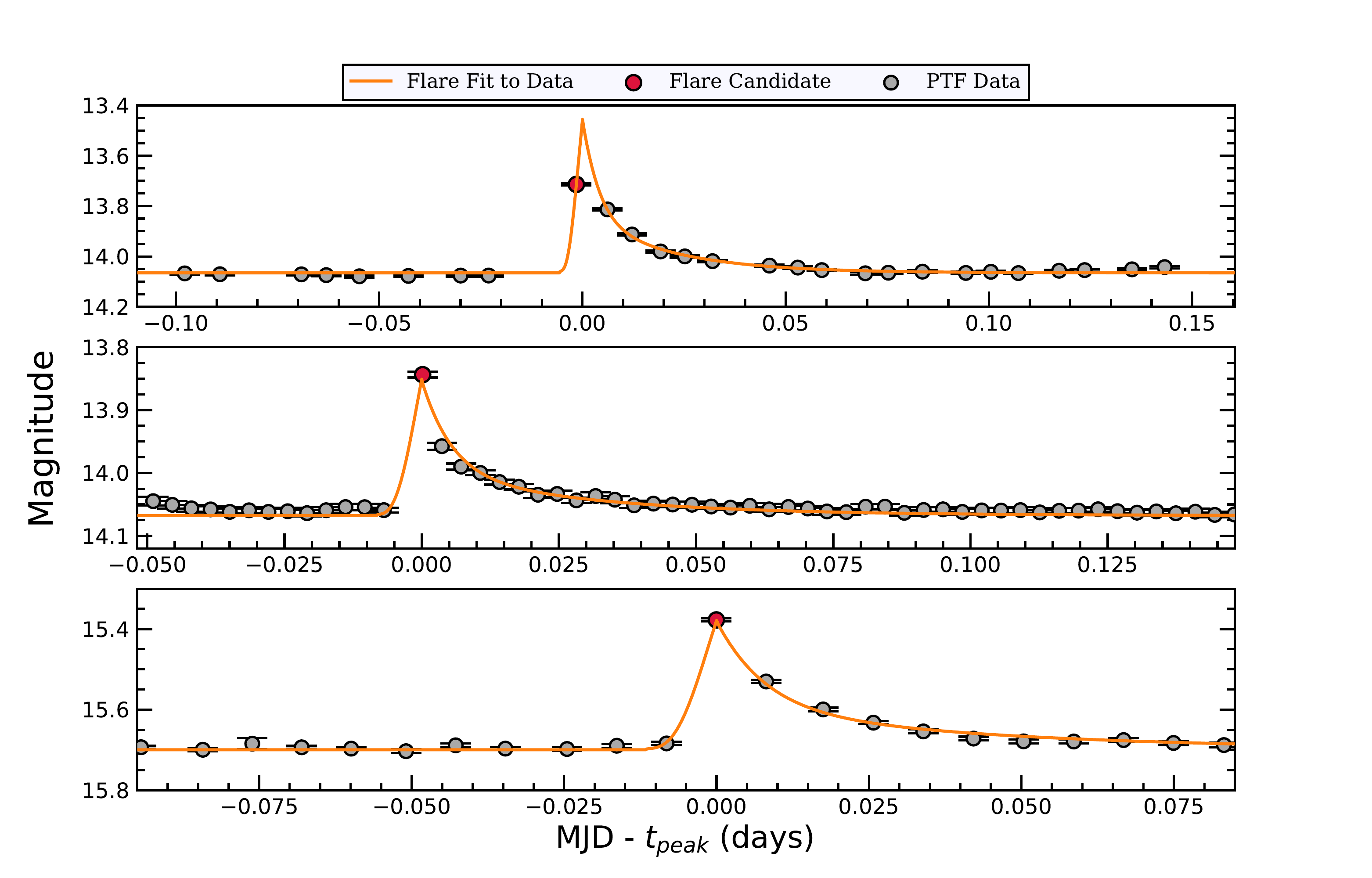}{0.95\textwidth}{}}
\caption{Three strong flare event candidates on M-type stars found in the TESS follow-up sample, identified using PyVAN. The orange line is the determined best flare template fit to the PTF observations (gray points), with the candidate flare peak observation in red. Though the middle and bottom events are fairly well constrained by observations, the top event allows a sizeable range of template parameters which correspond to approximately the same likelihood. Notably, peaks both before or after the candidate observation can result in reasonable fits for this target. The presented fit is the lowest amplitude event producing a comparable likelihood to the overall best-fit (see Section \ref{sec:lums}). \label{fig:tess_flares}}
\end{figure}

\begin{figure}
\gridline{\fig{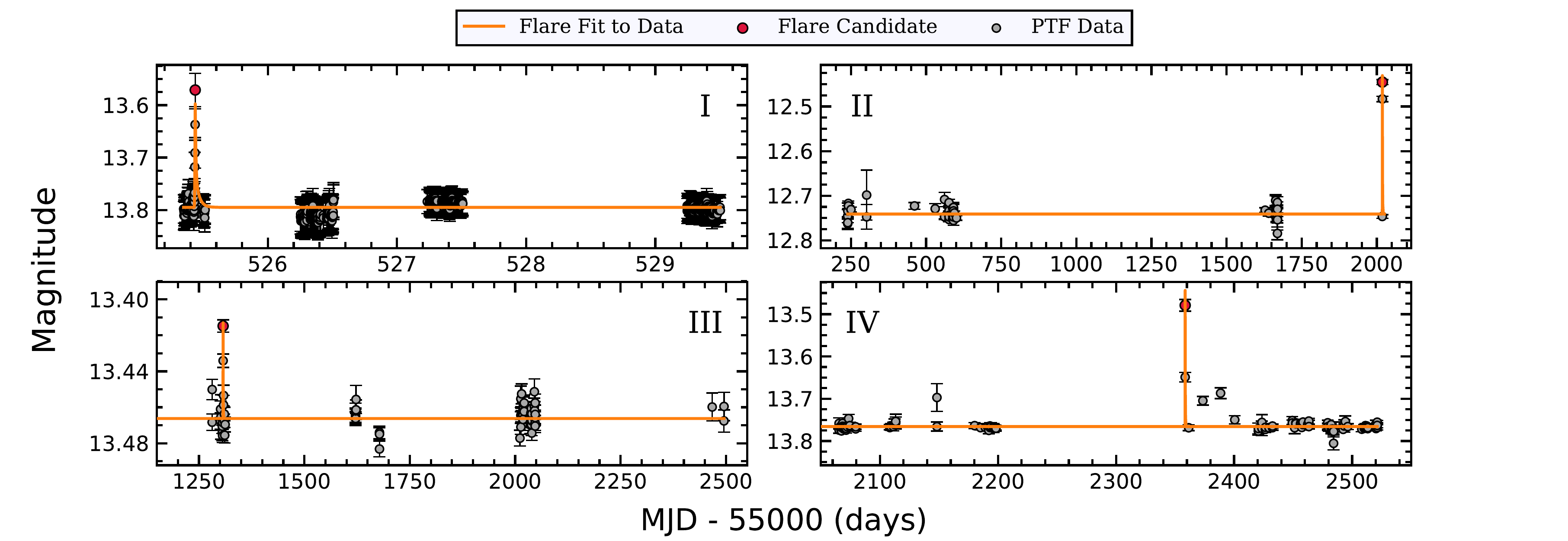}{0.95\textwidth}{}}
\caption{PTF light-curves for candidate flaring G-dwarfs with overlaid flare template fit (orange line). The number in the inner corner of each plot indicates the corresponding row in Table \ref{tab:flaringG}. \label{fig:Gtess}}
\end{figure}

Applying our techniques to the sample of targets for possible TESS follow-up results in an initial sample of 170 g-band candidates and 255 R-band candidates with $\Delta\ell$ values exceeding our thresholds (see Figure \ref{fig:recovered_tars} and Figure \ref{fig:tess_flares}). We approximate a spectral type for each passing target by converting PS1 colors to $B-V$ and $V-I_C$, again using color transformations from \citet{tonry2012}, and assigning spectral types for these colors based on corresponding values from \citet{pecaut2013}. Of the full sample, we select a small set of G-type candidates as targets of particular interest for follow-up with high-cadence TESS observations. In addition to requiring passing $\Delta\ell$ values, we require that the images resulting in each target’s candidate event is clear of ghosts, halos, or other contributors of possible flux enhancements. Further, since flares on G-type stars have lower contrast with their stellar flux on average (compared to M-type flares), and thus manifest as less statistically significant outliers in our photometry, we require corroboration of flare-like behavior with light-curves from The Catalina Surveys \citep{drake2009}. Specifically, we require at least a single clear outlier observation upon visual inspection of the Catalina Survey’s data. Ultimately, there are 4 targets meeting this criteria (see Table \ref{tab:flaringG} and Figure \ref{fig:Gtess}).

\begin{deluxetable}{ccccccc}
\tablewidth{0pt}
\tablecaption{Candidate G-type Stars Identified for TESS Follow-up}
\tablehead{\colhead{RA} & \colhead{Dec} & \colhead{$\Delta\ell_{fq}$} & \colhead{$\Delta\ell_{fr}$}  & \colhead{Est. Spectral Type} & \colhead{PTF Filter} & \colhead{$m_0$}}
\startdata
85.10256062 & -6.45354963 & 177.0 & 57.4 & G5V & R & 13.80 \\
130.0539418	& 21.11539871 & 95.5 & 29.1 & G6V & R & 12.74 \\
110.8485918 & 26.54241250 & 43.6 & 43.5 & G3V & R & 13.47 \\
190.6555490 & 29.03659331 & 93.3 & 78.3 & G1V & g & 13.77 \\
\enddata
\tablecomments{Selection of candidate G-type stars identified in PTF data for possible follow-up with TESS. These targets' $\Delta\ell_{fq}$ and $\Delta\ell_{fr}$ values place them well above the identified $1\%$ contamination thresholds (see Section \ref{sec:sims}). The process for spectral type estimation is explained in Section \ref{sec:tess}. $m_0$ refers to each target's fit quiescent magnitude in the given PTF filter.}\label{tab:flaringG}
\end{deluxetable}

\section{Future Applications}\label{sec:additional_apps}

PyVAN is well-suited for searches of flares (or other variables) in the data of a number of upcoming or recent missions. In the case of LSST data, for instance, PyVAN could serve as a complement for the ANTARES system, which does not currently accommodate flare detection \citep{narayan2018}.

Though designed for application to sparse ground-based data, PyVAN can also be applied to higher quality data of space-based surveys (e.g. TESS and Kepler). Applying the software as designed for PTF, stars with flare enhancements in excess of any star-spot modulation can be easily separated from those of common contaminants. Efficiency for this use could be improved by application of a detrending routine to candidate light-curves before flare fitting (e.g. to flatten spot modulation) and by tuning parameter bounds used in fitting to better utilize the significant constraints provided by such data.

\section{Summary}
We have presented PyVAN, a publicly available template optimization software especially suited for aiding in the recovery of the flaring status of stars from time-domain surveys. PyVAN uses non-linear optimization algorithms to fit light-curve templates of flaring, RR Lyrae, and quiet stars to candidate light-curves. Comparing the best odds that each template produced the observed data, PyVAN is able to isolate strong flare star candidates which are also unlikely to be explained by other astrophysical phenomena.

Applying PyVAN to simulated data of comparable signal-to-noise ratio and sampling to that of PTF, we have shown that the technique is capable of filtering out the most pervasive contaminants found in prior samples of flare stars, while recovering a safe population of flaring candidates. In these tests, PyVAN was able to identify 82\% of flare events that we determined to be recoverable. We further estimate from these results an overall flare star recovery rate of 1.5\% for data of similar quality. In application to a PTF sample of M-dwarfs from Stripe 82 and the Kepler field, we have demonstrated PyVAN's ability to produce a sample of flaring candidates that is consistent with previous literature. We have recovered the flaring status of 10 M-dwarf stars from the sample of 236 flaring M-dwarfs (4.2\%) reported in \citet{kowalski2009}, as well as 3 from a sub-sample of 158 flaring M-dwarfs (1.9\%) taken from the full sample reported in \citet{davenport2016}. For a combined recovery rate of around 3\%, these results are consistent with the speculation that the recovery rate approximated from testing on simulated data was reduced by a number of deficiencies in the simulated sample. Though rates of a few percent may seem to imply a low recovery efficiency, the stochastic nature of flare events combined with time-limited survey data should be expected to place a similarly small upper limit on the rate of recovery that is possible. Searching the light-curves of other M-dwarfs in Stripe 82, we recover an additional 245 flaring candidates for which flare events were not identified in the \citet{kowalski2009} sample. Applying our technique to a sample of PTF targets for possible TESS follow-up, we identify 4 candidate G-type flare stars. Finally, we note potential future applications of the PyVAN software to a number of recent and upcoming missions.

By utilizing PyVAN, or processes implementing similar techniques, samples of probable flaring stars can be produced which have significantly diminished rates of contamination. Mitigating current contamination helps to both improve the efficacy of flare population statistics and to maximize the value of observations with next generation observatories.

\acknowledgements
We thank our referee for providing feedback that helped to improve the content and clarity of this manuscript. We also acknowledge support from a Research Corporation Scialog grant.
\vspace{2mm}

This paper is based on observations obtained with the Samuel Oschin Telescope and the 60 inch Telescope at the Palomar Observatory as part of the Palomar Transient Factory project, a scientific collaboration between the California Institute of Technology, Columbia University, Las Cumbres Observatory, the Lawrence Berkeley National Laboratory, the National Energy Research Scientific Computing Center, the University of Oxford, and the Weizmann Institute of Science.

The Intermediate Palomar Transient Factory project is a scientific collaboration among the California Institute of Technology, Los Alamos National Laboratory, the University of Wisconsin, Milwaukee, the Oskar Klein Center, the Weizmann Institute of Science, the TANGO Program of the University System of Taiwan, and the Kavli Institute for the Physics and Mathematics of the Universe. 
\vspace{2mm}

The CSS survey is funded by the National Aeronautics and Space
Administration under Grant No. NNG05GF22G issued through the Science
Mission Directorate Near-Earth Objects Observations Program.  The CRTS
survey is supported by the U.S.~National Science Foundation under
grants AST-0909182 and AST-1313422.
\vspace{5mm}



\begin{thebibliography}{}

\bibitem[Abazajian et al.(2009)]{abazajian2009} Abazajian, K.N., Adelman-McCarthy, J.K., Agueros, M.A. et al. 2009, ApJS, 182, 543

\bibitem[Bellm et al.(2019)]{bellm2018} Bellm, E. C., Kulkarni, S. R., Graham, M. J., et al. 2019, PASP 131 018002

\bibitem[Bla{\v z}ko(1907)]{blazhko1907} Bla{\v z}ko, S.\ 1907, Astronomische Nachrichten, 175, 325

\bibitem[Borucki et al.(2010)]{borucki2010} Borucki, W.J., Koch, D., Basri, G. et al. 2010, Science, 327, 977

\bibitem[Dark Energy Survey Collaboration(2016)]{dark2016} Dark Energy Survey Collaboration, Abbott, T., Abdalla, F.B. et al. 2016, MNRAS, 460, 1270

\bibitem[Davenport(2016)]{davenport2016}
Davenport, J. R. A., 2016, ApJ, 829, 23
 
\bibitem[Davenport et al.(2014)]{davenport2014}
Davenport, J. R., Hawley, S. L., Hebb, L., Wisniewski, J. P., Kowalski, A. F. et al. 2014, ApJ, 797, 122

\bibitem[Debosscher et al.(2011)]{debosscher2011}
Debosscher, J., Blomme, J., Aerts, C., \& De Ridder, J. 2011, A\&A, 529, A89

\bibitem[Drake et al.(2009)]{drake2009}
Drake, A.~J., Djorgovski, S.~G., Mahabal, A., et al.\ 2009, Bulletin of the American Astronomical Society, 41, 470.07 

\bibitem[Fuhrmeister et al.(2011)]{fuhrmeister2011} Fuhrmeister, B., Lalitha, S., Poppengaeger, K., Rudolf, N., Liefke, C., Reiners, A., Schmitt, J.H.M.N., \& Ness, J.-U. 2011, A\&A, 534, 133

\bibitem[Gaia Collaboration et al.(2016)]{gaia}
Gaia Collaboration, T. Prusti, J.H.J. de Bruijne, A. Brown, A. Vallenari, C. Babusiaux, C.A.L. Bailer-Jones et al. (2016) The Gaia mission. A\&A 595, pp. A1.

\bibitem[Gaia Collaboration et al.(2018)]{gaiadr2}
Gaia Collaboration, A. G. A. Brown, A. Vallenari, T. Prusti, J. H. J. de Bruijne, C. Babusiaux, C. A. L. Bailer-Jones. (2018b) Gaia Data Release 2. Summary of the contents and survey properties. ArXiv e-prints.

\bibitem[Green et al.(2018)]{green2018} 
Green, G.~M., Schlafly, E.~F., Finkbeiner, D., et al.\ 2018, \mnras, 478, 651 

\bibitem[Hawley \& Pettersen(1991)]{hawley1991} Hawley, S.L. \& Pettersen, B.R. 1991, ApJ, 378, 725

\bibitem[Hawley et al.(2003)]{hawley2003} Hawley, S.~L., Allred, J.~C., Johns-Krull, C.~M., et al.\ 2003, \apj, 597, 535 

\bibitem[Hawley et al.(2014)]{hawley2014} 
Hawley, S.~L., Davenport, J.~R.~A., Kowalski, A.~F., et al.\ 2014, \apj, 797, 121 

\bibitem[Hilton(2011)]{hilton2011} Hilton, E.J. 2011, PhD thesis, University of Washington

\bibitem[Howard et al.(2018)]{howard2018} Howard, W.S., Tilley, M.A., Corbett, H. et al. 2018, ApJL, 860, 30

\bibitem[Ivezi{\'c} et al.(2008)]{ivezic2008} Ivezi{\'c}, {\v Z}., Kahn, S.~M., Tyson, J.~A., et al.\ 2008, arXiv:0805.2366

\bibitem[Juric(2012)]{juric2012}
Juric, M. 2012, LSD: Large Survey Database framework, Astrophysics Source Code Library

\bibitem[Kinemuchi et al.(2006)]{kinemuchi2006} Kinemuchi, K., Smith, H.~A., Wo{\'z}niak, P.~R., McKay, T.~A., \& ROTSE Collaboration 2006, \aj, 132, 1202 

\bibitem[Kirk et al.(2016)]{kirk2016}
Kirk, B., Conroy, K., Prša, A. et al. 2016, AJ, 151, 68

\bibitem[Kowalski et al.(2009)]{kowalski2009}
Kowalski, A. F., Hawley, S. L., Hilton, E. J., et al. 2009, AJ, 138, 633

\bibitem[Kowalski et al.(2010)]{kowalski2010} Kowalski, A.F., Hawley, S.L., Holtzman, J.A., Wisniewski, J.P., \& Hilton, E.J. 2010, ApJL, 714, 98

\bibitem[Kowalski et al.(2013)]{kowalski2013} Kowalski, A.F., Hawley, S.L., Wisniewski, J.P., Osten, R.A., Hilton, E.J., Holtzman, J.A., Schmidt, S.J., \& Davenport, J.R.A. 2013, ApJS, 207, 15

\bibitem[Kowalski et al.(2019)]{kowalski2018} Kowalski, A.~F., Wisniewski, J.~P., Hawley, S.~L., et al.\ 2019, \apj, 871, 167 

\bibitem[Lacy et al.(1976)]{lacy1976} Lacy, C.H., Moffett, T.J., \& Evans, D.S. 1976, ApJS, 30, 85

\bibitem[Law et al.(2009)]{law2009}
Law, N. M., Kulkarni, S. R. et al. 2009, PASP, 121, 1395

\bibitem[Law et al.(2015)]{law2015} Law, N., Fors, O., Ratzloff, J., et al. 2015, PASP, 127, 234

\bibitem[Lomb(1976)]{lomb1976} Lomb, N.~R.\ 1976, \apss, 39, 447 

\bibitem[Loyd et al.(2018)]{loyd2018a} Loyd, R.O. Parke, France, K., Youngblood, A. et al. 2018, ApJ, 867, 71

\bibitem[Loyd et al.(2018)]{loyd2018b} Loyd, R.O. Parke, Shkolnik, E.L., Schneider, A.C., Barman, T.S., Meadows, V.S., Pagano, I., \& Peacock, S. 2018, ApJ, 867, 70 

\bibitem[Maehara et al.(2012)]{maehara2012}
Maehara, H., Shibayama, T., et al. 2012, Nature, 485(7399), 478-481

\bibitem[Magnier et al.(2013)]{magnier2013}
Magnier, E. A., Schlafly, E. et al. 2013, The Astrophysical Journal Supplement Series, 205(2), 20

\bibitem[Moffett(1974)]{moffett1974} Moffett, T.J. 1974, ApJS, 273, 29

\bibitem[Narayan et al.(2018)]{narayan2018} Narayan, G., Zaidi, T., Soraisam, M.D. et al. 2018, ApJS, 236, 9

\bibitem[Nemec et al.(2013)]{nemec2013}
Nemec, J. M., Cohen, J. G., Ripepi, V., et al. 2013, ApJ, 773, 181

\bibitem[Newville et al.(2014)]{newville2014}
Newville, M., Stensitzki, T., Allen, D. B., \& Ingargiola, A. 2014, Zenodo, http://doi.org/10.5281/zenodo.11813

\bibitem[Nutzman \& Charbonneau(2008)]{nutzman2008} Nutzman, P. \& Charbonneau, D. 2008, PASP, 120, 317

\bibitem[Osten et al.(2012)]{osten2012} Osten, R.~A., Kowalski, A., Sahu, K., \& Hawley, S.~L.\ 2012, \apj, 754, 4 

\bibitem[Pecaut \& Mamajek(2013)]{pecaut2013} Pecaut, M.~J., \& Mamajek, E.~E.\ 2013, \apjs, 208, 9 

\bibitem[Pedregosa et al.(2011)]{pedregosa2011}
Pedregosa, F., Varoquaux, G. et al. 2011, Journal of Machine Learning Research, 12(Oct), 2825-2830

\bibitem[Pitkin et al.(2014)]{pitkin2014}
Pitkin, M., Williams, D., Fletcher, L., \& Grant, S. D. T. 2014, MNRAS, 445, 2268

\bibitem[Richards et al.(2011)]{richards2011} Richards, J.~W., Starr, D.~L., Butler, N.~R., et al.\ 2011, \apj, 733, 10 

\bibitem[Ricker et al.(2015)]{ricker2015}
Ricker, G. R., Winn, J. N. et al. 2015, Journal of Astronomical Telescopes, Instruments, and Systems, 1(1), 014003-014003

\bibitem[Scargle(1982)]{scargle1982} Scargle, J.~D.\ 1982, \apj, 263, 835 

\bibitem[Schmidt et al.(2018)]{schmidt2018} 
Schmidt, S.~J., Shappee, B.~J., van Saders, J.~L., et al.\ 2018, ApJ, submitted (arXiv:1809.04510)

\bibitem[Sesar et al.(2009)]{sesar2009}
Sesar, B., Ivezić, Ž., et al. 2009, ApJ, 708, 717

\bibitem[Segura et al.(2010)]{segura2010} Segura, A., Walkowicz, L.M., Meadows, V., Kasting, J., \& Hawley, S. 2010, Astrobiology, 10, 1089

\bibitem[Shappee et al.(2014)]{shappee2014} Shappee, B.J., Prieto, J.L., Grupe, D. et al. 2014, ApJ, 788, 48

\bibitem[Silverberg et al.(2016)]{silverberg2016}
Silverberg, S. M., Kowalski, A. F., Davenport, J. R., Wisniewski, J. P. et al. 2016, ApJ, 829, 129

\bibitem[Stetson(1996)]{stetson1996} Stetson, P.~B.\ 1996, \pasp, 108, 851

\bibitem[Storn \& Price(1997)]{storn1997}
Storn, R., \& Price, K. 1997, Journal of global optimization, 11(4), 341-359

\bibitem[Tonry et al.(2012)]{tonry2012} 
Tonry, J. L., Stubbs, C. W., Lykke, K. R., et al.\ 2012, \apj, 750, 99 

\bibitem[Vida \& Roettenbacher(2018)]{vida2018} 
Vida, K., \& Roettenbacher, R.~M.\ 2018, \aap, 616, A163

\bibitem[Walkowicz et al.(2011)]{walkowicz2011} Walkowicz, L.M., Basri, G., Batalha, N. et al. AJ, 141, 50

\bibitem[Yang et al.(2017)]{yang2017} Yang, H., Liu, J., Gao, Q., Fang, X., Guo, J., Zhang, Y., Hou, Y., Wang, Y., \& Cao, Z. 2017, ApJ, 849, 36

\bibitem[Zinn et al.(2017)]{zinn2017} Zinn, J.C., Kochanek, C.S., Kozlowski, S. et al. 2017, MNRAS, 468, 2189

\end{thebibliography}
\end{document}